\documentclass[prl,superscriptaddress,twocolumn,nopreprintnumbers, floatfix,preprintnumbers,altaffilletter]{revtex4}

\usepackage{graphicx,url,amssymb,amsmath,longtable,rotating,color,units,wasysym,subfigure,epsfig,multirow}

\usepackage{dcolumn}
\usepackage{bm}
\usepackage{hyperref}
\usepackage{txfonts}
\usepackage{ulem}

\usepackage[svgnames]{xcolor}

\hypersetup{
 colorlinks=true,
 allcolors=blue}

\newcommand{\be}{\begin{equation}}
\newcommand{\ee}{\end{equation}}
\newcommand{\bea}{\begin{eqnarray}}
\newcommand{\eea}{\end{eqnarray}}
\newcommand{\omgw}{\Omega_\mathrm{GW}}

\newcommand{\mrm}{\mathrm}
\newcommand{\mchirp}{\mathcal{M}_\mathrm{chirp}}

\begin{document}


\title{Postmerger: a new and dominant contribution to the
  gravitational-wave background from binary neutron stars}

\author{L\'eonard~Lehoucq}
\affiliation{Sorbonne Universit\'e, CNRS, UMR 7095, Institut
  d'Astrophysique de Paris (IAP), 98 bis boulevard Arago, 75014 Paris,
  France}

\author{Irina~Dvorkin}
\affiliation{Sorbonne Universit\'e, CNRS, UMR 7095, Institut
  d'Astrophysique de Paris (IAP), 98 bis boulevard Arago, 75014 Paris,
  France}
\affiliation{Institut Universitaire de France, Minist\`ere de
  l'Enseignement Sup\'erieur et de la Recherche, 1 rue Descartes, 75231
  Paris Cedex F-05, France}

\author{Luciano Rezzolla}
\affiliation{Institut f\"ur Theoretische Physik, Goethe Universit\"at,
  Max-von-Laue-Str. 1, 60438 Frankfurt am Main, Germany}
\affiliation{Frankfurt Institute for Advanced Studies,
  Ruth-Moufang-Str. 1, 60438 Frankfurt am Main, Germany}
\affiliation{School of Mathematics, Trinity College, Dublin 2, Ireland}


\begin{abstract}
The stochastic gravitational-wave background (SGWB) generated by the
inspiral and merger of binary neutron stars is traditionally modelled
assuming that the inspiral is promptly followed by the collapse of the
merger remnant to a rotating black hole. While this is reasonable for the
most massive binaries, it is not what is expected in general, as the
remnant may survive for up to hundreds of milliseconds and radiate an
amount of energy that is significantly larger than that lost during the
whole inspiral. To account for this additional contribution to the SGWB,
we consider a waveform model that includes both the inspiral and the
postmerger emission. We show for the first time that for a large set of
parameterized equations of state (EOSs) compatible with observational
constraints, there is considerable spectral power in the $1-2\,{\rm kHz}$
range, distinct from that associated with the inspiral and leading to a
dimensionless GW energy density $\Omega_{\rm GW} \simeq
10^{-10}-10^{-9}$. We discuss the enhanced detectability of the SGWB by
third-generation detectors such as the Einstein Telescope and Cosmic
Explorer, and show how it depends on the signal-to-noise ratio of
foreground binaries and on the remnant lifetime. Interestingly, even a
non-detection of the high-frequency part of the SGWB could provide
valuable constraints on the remnant lifetime, offering novel insights
into the postmerger dynamics and the EOS of nuclear matter.
\end{abstract}

\maketitle


{\it Introduction}--- The discovery of the binary neutron star (BNS)
merger GW170817 \cite{2017PhRvL.119p1101A} marked a major milestone in
multi-messenger astrophysics, as the observations of gravitational waves
(GW) by LIGO-Virgo~\cite{2015CQGra..32g4001L, 2015CQGra..32b4001A} was
followed by the detection of electromagnetic counterparts across multiple
wavelengths \cite{2017Sci...358.1556C, 2017ApJ...848L..14G,
  2017ApJ...848L..15S, 2018Natur.561..355M, 2019Sci...363..968G}. These
observations confirmed the connection between short gamma-ray bursts,
kilonovae and BNS mergers and provided important information on the
geometry of the jet launched after the merger as well as on the ejected
neutron-rich material. Joint analysis of electromagnetic and GW data led
to tight constraints of General Relativity \cite{2019PhRvL.123a1102A},
whereas the identification of the host galaxy allowed an independent
measurement of the Hubble parameter (see, e.g.,
\cite{2017Natur.551...85A}). Combined with the second, GW-only, BNS
detection~\cite{2020ApJ...892L...3A}, the local merger rate is
constrained to $10-1700\, {\rm Gpc}^{-3}\,{\rm yr}^{-2}$
\cite{the_ligo_scientific_collaboration_population_2022}. This
information is crucial for building population models and understanding
the formation of binary compact objects.

Observations of BNS mergers also provide a unique probe of the physics of
extremely dense matter and allow to study the equations of state (EOS) of
neutron star interiors. Since the EOS strongly affects the gravitational
waveform, it leaves an imprint in the tidal deformation that takes place
in the late inspiral stages and allows measuring the tidal
deformability~\cite{2015PhRvL.114p1103B, 2016PhRvL.116r1101H,
  2014MNRAS.437L..46R, 2019PhRvD..99b4029D,PhenomD_tidalv2}. In the case
of the GW170817 event, GW observations allowed to constrain the tidal
deformability parameter to $\tilde{\Lambda} \lesssim 700$, where
$\tilde{\Lambda}$ is the mass-weighted combination of the dimensionless
deformability parameters~\cite{PhysRevLett.121.161101,
  2019PhRvX...9a1001A}.

Additional and complementary information on the EOS can be extracted from
the postmerger signal emitted by the hypermassive neutron star (HMNS)
formed after the neutron stars have coalesced \cite{2017PhRvD..96l3012S,
  Rezzolla_bigReview, 2020ARNPS..70...95R}. In particular, a number of
recent studies have shown that the GW emission from the HMNS is
concentrated at specific frequencies, resulting in relatively narrow
spectral features~\cite{2011MNRAS.418..427S, 2012PhRvL.108a1101B,
  2014PhRvL.113i1104T, 2015PhRvD..91l4056B, 2016PhRvD..93l4051R,
  2015PhRvD..91f4001T, 2024ApJ...960...86T}. The postmerger signal has
not been detected so far, as the relevant frequencies are above those
observable with current ground-based detectors. However, future
third-generation (3G) detectors, such as the Einstein Telescope (ET)
\cite{2020JCAP...03..050M} and the Cosmic Explorer (CE)
\cite{2021arXiv210909882E} will be able to detect and characterize the
postmerger signal for sufficiently nearby systems.

An alternative method to study populations of compact binaries is by
observing the stochastic GW background (SGWB) created by the incoherent
superposition of unresolved sources \cite{2011RAA....11..369R,
  christensen_stochastic_2019}. Different types of sources can contribute
to the SGWB, including compact binaries, core-collapse supernovae and
isolated spinning neutron stars \cite[e.g.][]{2005PhRvD..72h4001B,
  regimbau_astrophysical_2008, 2015PhRvD..92f3005C}. Observing the SGWB
will allow probing high-redshift populations, which are otherwise not
accessible to ground-based detectors, although various methods are being
developed to extract the SGWB from the detector noise in data from
Advanced LIGO, Advanced Virgo and Kagra (LVK)
\cite{regimbau_quest_2022,2022Galax..10...34R}. Current upper limits for
ground-based detectors are $\Omega_{\rm{GW}}(f)\leq 3.4\times 10^{-9}$ at
$f=25$ Hz, where $\Omega_{\rm{GW}}$ is the dimensionless energy density
of the SGWB \cite{PhysRevLett.119.029901, PhysRevLett.120.091101,
  PhysRevD.100.061101, PhysRevD.104.022004, PhysRevD.104.022005}. These
upper limits contain valuable information on high-redshift populations of
compact binaries, complementary to that of individual detections
\cite[e.g.][]{2015A&A...574A..58K, dvorkin_metallicity-constrained_2016,
  Nakazato_2016, perigois_startrack_2021, 2019PhRvD.100f3004C,
  2020ApJ...896L..32C, perigois_gravitational_2022, LL2023}. For the LVK
detectors, the dominant contribution to the SGWB is expected to be that
of binary black holes. However, this will not be the case for ET and CE,
since almost all the binary black hole mergers will be detected
individually and subtracted from the data. The remaining SGWB will thus
contain information on the unresolved BNS mergers.

In the case of the postmerger emission from a population of BNS mergers,
we expect that if HMNS are formed, the characteristic GW frequencies will
be enhanced in the SGWB due to the large number of unresolved sources
contributing to the signal. This opens up the possibility to detect this
EOS-dependent feature and to obtain complementary information on the
postmerger stage. The goal of this Letter is to model the SGWB from a
population of BNS, including the postmerger stage, and to highlight the
existence of a so-far neglected additional and dominant component that
can be detected with ET and CE.

{\it BNS inspiral and postmerger waveforms}--- To construct a hybrid
waveform, we combine in the frequency domain the analytic prescriptions
referring to the inspiral and the postmerger. This combination takes
place at a frequency $f_{\rm tr}$, where we move from the inspiral to the
postmerger description (see End Matter for details). This frequency must
be significantly smaller than the main frequencies appearing in the power
spectral density (PSD) of the postmerger signal, i.e., $f_1$ and $f_2$,
where $f_1 < f_2$, so that no power from the postmerger phase is
missed. Hereafter, for simplicity we set $f_{\rm tr} = f_1/2$, but the
overall results depend only weakly on this assumption. The PSD in
  the $f_1$ frequency is significantly suppressed in the case on
  unequal-mass binaries but is here included for consistency with the use
  of a population of neutron-star binaries with equal-mass components. In
  practice, because the corresponding power is about one order of
  magnitude smaller than that in the $f_2$ peak, its impact in shaping
  the high-frequency component of the SGWB is very small.

{\it The equation of state}--- In view of the uncertainties in the EOS
and their effect on the waveform, we adopt an agnostic approach in this
work. We use the findings of Refs.~\cite{Rezzolla_EOS_1, Rezzolla_EOS_2,
  Rezzolla_EOS_3, Rezzolla_EOS_4}, who constructed more than $10^7$ EOSs
consistent with constraints from fundamental physics and astronomical
observations of neutron stars. Leveraging on this extensive dataset, it
was possible to derive informative constraints on the relationship
between the chirp mass $\mchirp$ and tidal deformability $\Lambda$ in
terms of a probability density function (PDF) for these two quantities.
A-priori, a single EOS does not correspond to a unique line in the
$\mchirp$-$\Lambda$ plane because the same chirp mass may result from
many combinations of BNS masses. However, when restricting the analysis
to symmetric binaries, as we do throughout the rest of this work, this
ambiguity disappears. In this case, for a given EOS, each chirp mass is
uniquely associated to a single value of $\Lambda$. Therefore, extracting
an EOS from a given distribution is equivalent to drawing a single value
of $\Lambda$ from the PDF at each chirp mass. Similarly, extracting a
certain value of $\Lambda$ from the allowed range is equivalent to
singling-out a specific EOS.

Although the PDF found in~Ref.~\cite{Rezzolla_EOS_1} for a given chirp
mass is not exactly a Gaussian, it is so to a first approximation [see
  Fig.~(4) in~\cite{Rezzolla_EOS_1}]. Hence, when assuming an underlining
Gaussian distribution, we can relate the tidal deformability and the
chirp mass (in solar masses) via a simple power-law of the type $
\Lambda_p = \alpha + \beta \mathcal{M}^{\gamma}_\mathrm{chirp} $ where
$\Lambda_p$ is normally distributed between the minimum and maximum
values given by~\cite{Rezzolla_EOS_1} $\Lambda_{\rm min\,(max)} = a +
b\,\mathcal{M}_{\rm chirp}^{c}$ with $a=-50\,(-20)$, $b=500\,(1800)$, and
$c=-4.5\,(-5.0)$. We here consider $\gamma=-3$, but we have checked that
varying $\gamma$ leads to very little differences in SGWB, as the mass
distribution of BNS has a clear main peak. In this way, we can build a
set of fictitious $10^3$ EOSs such that the corresponding tidal
deformability is normally distributed.

Once an EOS is set, we can compute the inspiral waveform. For the
postmerger waveform, we need to know the characteristic frequencies $f_1$
and $f_2$ given the selected EOS and hence tidal deformability. Luckily,
this is rather straightforward to do since universal relations have been
found that provide analytic fits (see, e.g., \cite{2012PhRvL.108a1101B,
  2014PhRvL.113i1104T, 2015PhRvD..91l4056B, 2016PhRvD..93l4051R}). Here,
we employ the expressions for $f_1$ and $f_2$ proposed in
Refs.~\cite{2016PhRvD..93l4051R, 2024ApJ...960...86T} (and also employed
in~\cite{Rezzolla_main}). We note that being only ``quasi''-universal
relations, these expressions come with some intrinsic uncertainties as
estimated from the numerical simulations and the corresponding fits (see
discussion in~\cite{2024ApJ...960...86T}). These uncertainties, however,
do not play an important role here as we are interested in a SGWB that is
naturally broad given the much larger uncertainties in the EOS.

{\it Stochastic gravitational-wave background}--- The energy density in
GWs per logarithmic unit of frequency and in units of the critical
density of the Universe, $\rho_\mathrm{c} = 3H_0/8\pi G$, is given by
$\omgw := ({1}/{\rho_c}) d\rho_\mathrm{GW}/{d}\ln{f} =
({f}/{c\,\rho_\mathrm{c}})F(f)$, where $\rho_\mathrm{GW}$ is the GW
energy density and $F(f)$ is the energy flux. When considering a
population of $N$ sources, the energy flux per unit frequency is given
by~\cite{1999PhRvD..59j2001A}: $F(f) := ({\pi c^3\,f^2}/{2G\,T})
\sum_{i=1}^{N}[|\Tilde{h}_i^+(f)|^2+|\Tilde{h}_i^\times(f)|^2]$, where
the index $i$ sums over the $N$ sources and $\Tilde{h}_i^+(f)$ and
$\Tilde{h}_i^\times(f)$ are the Fourier transforms of the two GW
polarization modes. The number of sources is taken to be proportional to
the total observation time $T$. We have checked that our calculation
converges at $T=1$ year and will use this value in the following (note
that the background is stationary).

In order to construct our BNS catalogue, we follow the
\texttt{baseline\_delays} model from Ref.~\cite{LL2023} and, more
specifically, we assume all binaries are symmetric in mass and follow a
Gaussian mass distribution with a mean of $1.33\,M_\odot$ and a standard
deviation of $0.09\,M_\odot$. The BNS-merger rate is described by the
star-formation rate from Ref.~\cite{vangioni_impact_2015} convolved with
a distribution of delay times $t_\mrm{d}$ that follows a power-law
distribution $P(t_\mrm{d})\propto
t_\mrm{d}^{-1}$~\cite{2018MNRAS.474.2937C}. Using this model, we
construct a catalogue containing about $N=3.8\times~10^5$ BNS that merge
in the entire Universe during one year, consistent with other studies
\cite{2022ApJ...941..208I} (see Ref.~\cite{LL2023}).

As we focus on the background, we must consider the flux contributions
from non-detectable sources only. We assume that individual resolvable
sources can be identified when their signal-to-noise ratio (SNR) exceeds
a threshold $\mrm{SNR}_\mrm{thr}=12$ ~\cite{2015PhRvD..92f3002M,
  2022ApJ...941..208I}, so that the contribution from these resolvable
sources is removed by subtracting each individual signal, and we use the
\texttt{GWFish} package~\cite{GWFish} to compute the SNR. In this
initial study we assume that the subtraction is noiseless, and leave more
realistic approaches for future work~\cite{background_subtraction,
  2017PhRvL.118o1105R, 2020PhRvD.102b4051S, Pop+omegaerr}.

\begin{figure}
    \centering
    \resizebox{\hsize}{!}{\includegraphics{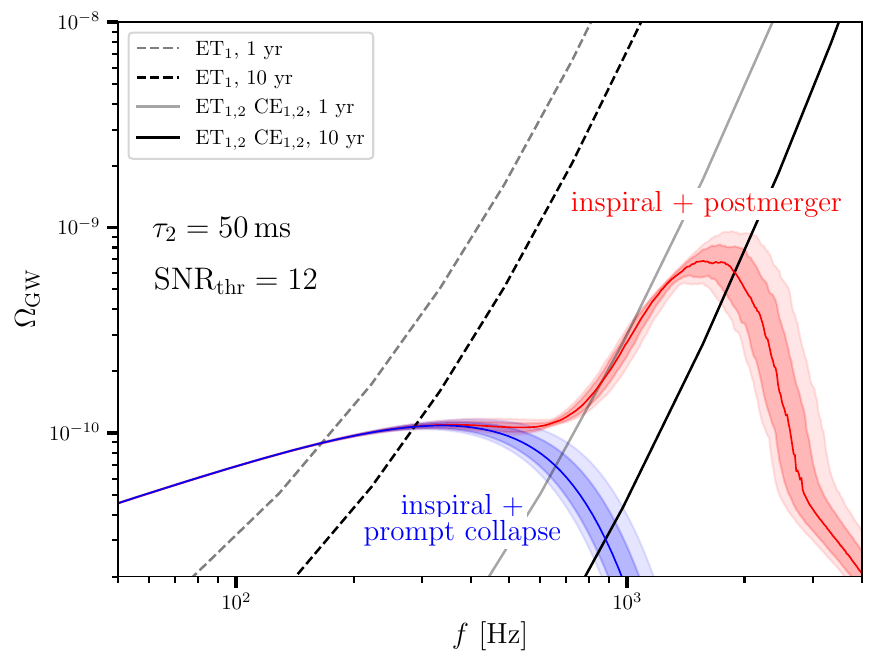}}
    \caption{Energy density of the SGWB from BNS mergers, corresponding
      to either the IPC (blue-shaded areas) or the IPM (red-shaded areas)
      scenarios. Shown with different transparencies are the $68$ and
      $95$ percentile contours, while the solid lines report the
      medians. Also outlined are the power-integrated sensitivity curves
      of a single ET (dashed lines) and of a network composed of an ET
      detector and of two CE detectors (solid lines) relative to $1$
      (grey lines) and $10$ (black lines) years of integration,
      respectively.}
    \label{fig:main+sensi}
\end{figure}

{\it Results}--- Figure~\ref{fig:main+sensi} shows the SGWB from BNS
mergers computed with our two descriptions for the modelling of the
signal. More specifically, in the \textit{inspiral+prompt collapse}
model, hereafter ``IPC'' (blue-shaded areas), which is the one adopted so
far in describing the SGWB from BNS mergers, the inspiral is assumed to
be followed by a prompt collapse to a BH of the merger remnant. On the
other hand, in the novel \textit{inspiral+postmerger} model, hereafter
``IPM'', proposed here (red-shaded areas), the signal includes the
postmerger emission from a metastable HMNS that eventually collapses into
a BH. For both models, we use ${\rm SNR}_\mrm{thr}=12$, and the
background was calculated using $10^3$ randomly sampled EOSs using the
methodology described above. The shaded regions correspond to the $68$
and $95$ percentiles of the resulting distribution, respectively, while
the solid lines indicate the median values in the distributions.

\begin{figure}
    \centering
    \resizebox{\hsize}{!}{\includegraphics{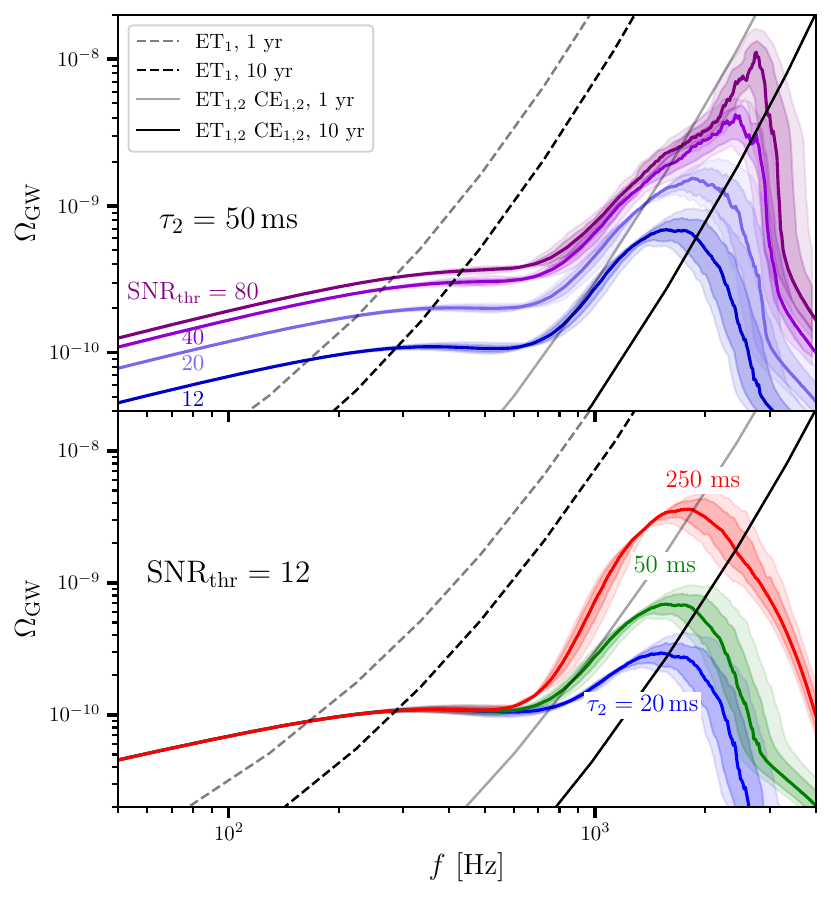}}
    \caption{\textit{Top panel:} The same as in Fig.~\ref{fig:main+sensi}
      but where different colours show how $\Omega_{\rm GW}$ changes when
      assuming threshold SNR values of ${\rm SNR_{thr}}= 12, 20, 40$, and
      $80$; note that $\tau_2=50\,{\rm ms}$ in all cases. \textit{Bottom
        panel:} The same as in the top but when considering different
      values for the decay timescale $\tau_{2}= 20, 50$ and $250\,{\rm
        ms}$ (blue, green, and red, respectively). Note that ${\rm
        SNR_{thr}}= 12$ in all cases.}
    \label{fig:main+SNRthr}
\end{figure}

As shown in Fig.~\ref{fig:main+sensi}, both models coincide at low
frequencies (i.e., below $200\,{\rm Hz}$), which is expected since their
inspiral parts are identical. They begin to diverge at approximately
$350\,{\rm Hz}$. The power in the IPC model rapidly decreases as
expected, since it does not include significant postmerger emission. In
contrast, the IPM model exhibits a peak followed by a gradually declining
plateau, and then experiences a rapid rise to a broad peak near
$1.7\,{\rm kHz}$, which is about six times higher than the $350\,{\rm Hz}$
peak in $99\%$ of the EOSs considered. This peak is broad because of the
redshift distribution of the sources and the intrinsic differences in the
$10^3$ EOSs considered. We note that the high-frequency peak arises
solely from the postmerger contribution, while the low-frequency peak is
primarily influenced by the inspiral phase, as indicated by the overlap
of the blue and red distributions. On the other hand, the width of the
SGWB peak is due mostly to the redshift distribution of the sources,
while the uncertainty in the EOSs contained in our sample is what leads
to the variance shown with the shaded areas.

Black solid and dashed lines in Fig.~\ref{fig:main+sensi} represent the
power-law integrated sensitivity curves of two 3G detector
configurations: a single ET configuration (dashed lines) and an
admittedly optimistic configuration consisting of $2$ ET detectors, as
well as two CE detectors (solid line), such type of network can
drastically improve the background sensitivity~\cite{network_sensi_main,
  CE, Maggiore2024}. Sensitivity curves for both scenarios are shown
either for one-year integration time (grey solid lines) or for ten years
of observations (black lines) and with an SNR for SGWB detection of $\rho
= 1$.

The results in Fig.~\ref{fig:main+sensi} clearly indicate that even after
ten years, a single 3G detector will not be able to reach the peaks in
the SGWB arising from the inspiral or from the postmerger. However, the
network setup could detect the low-frequency peak with just one year of
data and potentially the high-frequency peak with ten years of
observation. Note that the improvement in sensitivity between the two
scenarios is less pronounced than expected, primarily because ET and CE
are geographically distant, leading to a suppressed overlap-reduction
function at high frequencies, thereby limiting the potential gains.

We have so far assumed that all the individually detected sources are
subtracted noiselessly from the data, leaving out the clear SGWB from
unresolved BNS. In practice, however, this subtraction is challenging,
especially for low-SNR sources, whose parameters are recovered with large
uncertainties. To assess the impact that this subtraction has on the
estimated SGWB, we consider scenarios in which the individual detection
threshold is increased, which is equivalent to tagging as detected only
the loudest sources. In this case, we expect better source subtraction,
leading to less noise in the SGWB, which will also contain more sources.

We explore different detection thresholds in the left panel of
Fig.~\ref{fig:main+SNRthr}, where we compare our default value
$\mathrm{SNR}_{\mathrm{thr}} = 12$ with higher values of $20, 40$, and
$80$. As expected, the amplitude of the SGWB increases with
$\mathrm{SNR}_{\mathrm{thr}}$, which is encouraging as the background
approaches the network sensitivity. Indeed, as we increase
$\mathrm{SNR}_{\mathrm{thr}}$, fewer sources are identified as individual
detections, and instead contribute to the background. At the same time,
as $\mathrm{SNR}_{\mathrm{thr}}$ is increased, the positions of both the
low- and high-frequency peaks shift to higher frequencies. This shift
occurs because the background increasingly includes brighter sources,
which are typically closer and less redshifted. Therefore, the peaks move
to larger frequencies, where the detector sensitivity is lower, so that
the actual detection of the SGWB at these frequencies will remain
challenging despite the decrease in the background noise.

An important aspect of our analysis that is not firmly set is the
timescale $\tau_{2}$ for the damping of the $f_2$ frequency. This
timescale is clearly dependent on the EOS considered, but also on the
microphysical description of the postmerger evolution, where magnetic
fields, neutrino emission and ejection of matter all have an impact on
$\tau_{2}$. Given this enormous uncertainty, we can only assume that
$\tau_{2} \lesssim \tau_{\rm rem} \lesssim 1\,{\rm s}$, where $\tau_{\rm
  rem}$ is the remnant lifetime for which we have some constraints for
the only merger for which we have detected an electromagnetic counterpart
(in the case of the GW170817 event the merger lifetime has been estimated
to be of about one second~\cite{Gill2019, Murguia-Berthier2020}). Hence,
we report in the right panel of Fig.~\ref{fig:main+SNRthr} the SGWB for
the IPM model with a threshold $\mathrm{SNR}_{\mathrm{thr}} = 12$ and for
three different cases for the decay time, namely, $\tau_{2} = 20,\,50$,
and $250\,{\rm ms}$.

As expected, the PSD in the high-frequency peak of the postmerger signal
increases linearly with the increase of $\tau_{2}$, thus considerably
increasing the potential of detection of this part of the SGWB. At the
same time, integration times longer than one year are needed to detect
the high-frequency portion of the SGWB related to the postmerger.
Interestingly, however, the increase in the decay time also leads to an
increased power at lower frequencies, i.e., $500 \lesssim f \lesssim
1000\,{\rm Hz}$, so that the detectability of the high-frequency portion
of the SGWB is possible even if the maximum of the postmerger PSD is
still below sensitivity. In turn, the results shown in the right panel of
Fig.~\ref{fig:main+SNRthr} also reveal that the detection of the
high-frequency portion of the SGWB can provide precious information on
the lifetime of the remnant and hence on the EOS. It is unlikely that
stringent constraints on the EOS can be drawn from this measurement,
which however will serve as an important consistency check. We will
explore the implications of these results and how they can be used to
infer the nuclear-matter EOS in a future work.

{\it Conclusion}--- The SGWB generated by the inspiral and merger of
binary neutron stars has so far been modelled assuming that the inspiral
is promptly followed by the collapse of the merger remnant to a rotating
black hole. While this is reasonable for very massive binaries, it is not
what is expected to take place for systems having masses around the peak
of the mass distribution of neutron-star binaries. Hence, the most likely
outcome of the merger is that of producing a remnant that will survive
for hundreds of milliseconds before collapse. During its short existence,
the remnant can radiate enormous amounts of energy in GWs and, indeed,
the radiated GW energy is 3-4 times larger than that emitted over the
long inspiral.

These considerations have motivated us to include also the postmerger
contribution to the PSD and hence calculate how the SGWB is modified by
this important and dominant contribution. To this scope, we have
implemented an analytic model of the postmerger GW signal that has been
shown to provide a reasonably accurate description of the signal computed
by numerical-relativity simulations~\cite{Rezzolla_main}. The analytic
waveform is dominated by the fundamental frequency of oscillation of the
remnant $f_2/2$, that provides a narrow-banded emission around $f_2
\simeq 1-2\,{\rm kHz}$. Quasi-universal relations have been employed to
relate $f_2$ with the tidal deformability and the latter with the chirp
mass of the binary. Finally, the analytic description of the GW emission
and the parametric construction of the EOSs have then been combined with
a model for the population of equal-mass BNS binaries~\cite{LL2023}. In
this way, the postmerger contribution depends mostly on only two
parameters: the decay timescale of the GW at the $f_2$ frequency, i.e.,
$\tau_2$, and the threshold value of the SNR at which a BNS can be
resolved out of the background, i.e., ${\rm SNR_{thr}}$. Because both
$\tau_2$ and ${\rm SNR_{thr}}$ are poorly known, in our analysis they are
varied across the reasonable intervals in which they are expected to
range.

Using this setup, we were able to compute the high-frequency postmerger
contribution to the SGWB from about $\sim 3.8\times 10^{5}$ binaries
(i.e., those that are expected to merge in the whole universe in one
year~\cite{LL2023}), finding that this indeed represents a new and
dominant contribution. While being broad, this contribution peaks at
$\sim 1-2\,{\rm kHz}$ and provides a contribution to the normalised GW
energy density $\Omega_{\rm{GW}}\simeq 10^{-10}-10^{-9}$, hence much
larger than what is expected when considering scenarios with the inspiral
and collapse only. Clearly, being a background signal, the detectability
of this contribution will also depend on the integration time employed
and our analysis has shown that long integration times and a network of
3G detectors are likely to be needed for the detection. However, such a
network setup could detect the peak of the high-frequency component in
less than one year if $\tau_2 \gtrsim 100\,{\rm ms}$. More importantly,
the detection of the high-frequency portion of the SGWB can provide
information that is either complementary or supplementary to that
  of the foreground signals. More specifically, the detection of the
  high-frequency component would constrain the typical lifetime of the
  remnant and since the latter depends on the EOS, it would provide
  complementary information on the EOS. On the other hand, when
  considering the reconstruction of the BNS population and its evolution
  with redshift, the detection of the high-frequency component -- and
  specifically the location of the maximum and its broadness -- will tell
  us about the whole population of binaries beyond what can be deduced
  from the loudest foreground sources. Overall, the results presented in
this first analysis show that the postmerger contribution to the SGWB
cannot be ignored as done so far and that it contains crucial information
on the HMNS accessible with 3G detectors.

Notwithstanding the innovative nature of the procedure presented here,
our approach can be improved in a number of ways. First, by considering
that the decay time is not a simple mass dependant power-law, but
actually a function of the EOS. Second, by extending the analysis to
binaries that are not symmetric and hence introduce a degeneracy in the
modelling of the chirp mass. Finally, by considering a more careful
analysis of the foreground subtraction, which has been here treated
simply in terms of the threshold SNR. We will explore these refinements
of the model in future work.

\bigskip
{\it Acknowledgements}--- This research is supported by the \'Ecole
Normale Sup\'erieure Paris-Saclay CDSN PhD grant. This research is also
supported in part by the ERC Advanced Grant ``JETSET: Launching,
propagation and emission of relativistic jets from binary mergers and
across mass scales'' (grant No. 884631), by the Deutsche
Forschungsgemeinschaft (DFG, German Research Foundation) through the
CRC-TR 211 ``Strong-interaction matter under extreme conditions'' --
project number 315477589 -- TRR 211, and by the State of Hesse within the
Research Cluster ELEMENTS (Project ID 500/10.006). LR acknowledges the
Walter Greiner Gesellschaft zur F\"orderung der physikalischen
Grundlagenforschung e.V. through the Carl W. Fueck Laureatus Chair.

\appendix

\section*{End Matter}

\begin{figure}
    \centering \resizebox{\hsize}{!}{\includegraphics{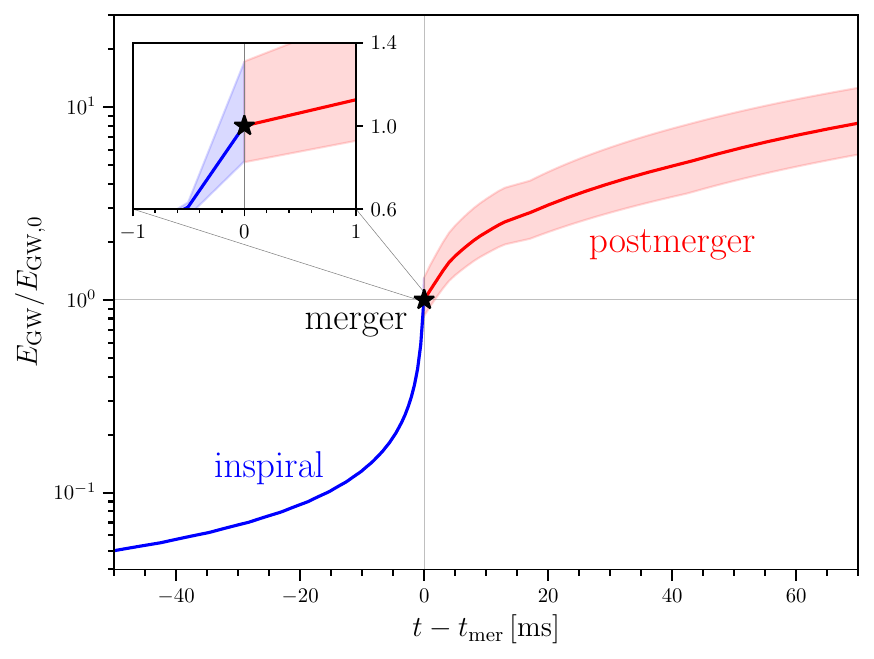}}
    \caption{Evolution of the radiated GW energy when normalized to the
      at the merger. Shown with blue- and red-shaded areas are the
      corresponding values during the inspiral (blue) and the postmerger.
      The variance in the shaded regions marks the $99\%$ percentile.}
    \label{fig:EGW}
\end{figure}

{\it On the energy radiated in GWs}--- A major focus of our work is to
extend the contribution of the BNS merger signal to the SGWB by
considering not only the inspiral part but also the postmerger
emission. To appreciate simply why this is important and hence the
relevance of our findings, we show in Fig.~\ref{fig:EGW} the total GW
energy radiated by a population of BNS when integrated over the entire
frequency range ($E_\mathrm{GW}$) and normalized by the total energy
considering only the inspiral phase ($E_{\mathrm{GW},0}$), as a function
of the time before and after the merger $t-t_{\rm mer}$. Near the merger
($t-t_{\rm mer} \lesssim 0$) the energy radiated increases very rapidly,
gaining an order of magnitude in the last $10\,{\rm ms}$ before the
merger. Following the merger ($t-t_{\rm mer} > 0$), the energy radiated
continues to grow at a slower rate but more intensely, so that it can
reach values much larger than those lost during the whole
inspiral. Stated differently, BNS mergers are copious emitters of GWs
during the inspiral, but it is the postmerger that provides the largest
contribution to the radiated energy.

Of course, the exact amount of energy radiated will depend sensitively on
the EOS describing the matter of which the BNS is composed of and which
is essentially unknown. Instead of concentrating on a few specific EOSs,
it is more convenient to consider a large ensemble of agnostic EOSs
satisfying astrophysical, GW, or elementary-particle
constraints~\cite{Annala2017, Most2018, Greif2019, Rezzolla_EOS_1}. This
is what is represented with the blue-shaded band in Fig.~\ref{fig:EGW},
which reports the 99th percentile contour of an ensemble of $10^3$ EOSs
generated using our EOS-agnostic approach. Similarly, the solid blue line
reports the median of the distribution at each time in the evolution. In
this way, it is straightforward to appreciate that since most of the
energy is lost after the merger, it is important to include also the
postmerger emission when calculating the BNS GW background (similar
considerations apply when considering the radiated angular
momentum~\cite{ecker:2024b, bernuzzi2015b}).

{\it Modelling the BNS waveforms}--- To model the inspiral part of the
GW waveform, we use the \texttt{IMRPhenomD\_NRTidalv2} model
\cite{2019PhRvD..99b4029D, PhenomD_tidalv2} from the LALSuite simulation
package~\cite{2020ascl.soft12021L}. This model extends the
\texttt{IMRPhenomD} waveform~\cite{PhenomD1,PhenomD2}, designed for
binary black hole systems, by incorporating EOS-dependent tidal
effects. Specifically, the template uses the dimensionless tidal
deformability parameter $\Lambda:=G^4 \lambda /c^5 M^5$, where $M$ is the
stellar mass and the tidal deformability parameter $\lambda$ is defined
as the ratio between the induced quadrupole and the external
gravitational field ($G$ and $c$ are the gravitational constant and the
speed of light, respectively). The model assumes that the merger remnant
promptly collapses to a black hole, thus leading to very little power in
the postmerger phase~\cite{PhenomD_tidalv2}. As commented above, this is
a very drastic assumption and applicable only to a rather restricted set
of binaries with total gravitational mass that is roughly given by
$M_{\rm tot} \gtrsim 1.45\,M_{_{\rm TOV}}\, \sim 3.38\,
M_{\odot}$~\cite{Koeppel2019}, where $M_{_{\rm TOV}}$ is the maximum
nonrotating gravitational mass for a given EOS and the estimate comes
from assuming a maximum mass $M_{_{\rm TOV}} \sim
2.33\,M_{\odot}$~\cite{Margalit2017, Rezzolla2017} (see
also~\cite{Bauswein2013, Tootle2021, Koelsch2021, Kashyap:2021wzs} for a
discussion on the threshold mass in BNS mergers).

Numerical-relativity simulations suggest that BNS mergers typically
produce a metastable HMNS, which persists for several tens of
milliseconds before ultimately collapsing into a spinning black
hole~\cite{Rezzolla_bigReview, Paschalidis2016}. Recent studies showed
that the GW signal emitted by this HMNS exhibits several prominent peaks
in the frequency domain~\cite{2011MNRAS.418..427S, 2012PhRvL.108a1101B,
  2015PhRvD..91l4056B, 2016PhRvD..93l4051R,2014PhRvL.113i1104T,
  2015PhRvD..91f4001T,2024ApJ...960...86T}. Here, we use the results
of~Ref.~\cite{Rezzolla_main}, which presents analytical fits for the
postmerger GW signal based on an extensive catalogue of time-domain
numerical waveforms. These fits express the time-domain strain as a sum
of damped sinusoids with a time-evolving instantaneous frequency as
$h_+(t) = h_0 \, \mathrm{e}^{-t/\tau_1} [ \mathrm{sin}(2\pi f_1 t) +
  \mathrm{sin}(2\pi (f_1 - f_{1\epsilon}) t) + \mathrm{sin}(2\pi (f_1 +
  f_{1\epsilon}) t) ] + \mathrm{e}^{-t/\tau_2} \, \mathrm{sin} ( 2\pi f_2
t + 2\pi \gamma_2 t^2 + 2\pi \xi_2 t^3 + \pi \beta_2 )$, where $h_0 =
G\mchirp/(2Dc^2)$ with $D$ the distance to the source, $t=0$ is the
merger time, the frequencies $f_1$ and $f_2$ represent the dominant
postmerger GW modes with $f_2 > f_1$, $\tau_1$ and $\tau_2$ are the decay
times of each mode, and $f_{1\epsilon}$, $\beta_2$, $\gamma_2$ and
$\xi_2$ are constants~\cite{2011MNRAS.418..427S, Takami2014,
  2016PhRvD..93l4051R}. The first peak, $f_1$, is short-lived (with a
decay time $\tau_1$ of the order of a few ms) and most probably
originates from damped collisions of the two stellar cores immediately
after the merger~\cite{Takami2015}. The second peak, $f_2$, corresponds
to the spin frequency of the deformed HMNS and has a decay time $\tau_2$
of the order of several tens of milliseconds~\cite{Takami2014,
  Bauswein2015}. However, the expected values, and mass dependencies of
both $\tau_1$ and $\tau_2$ are still rather uncertain as they sensitively
depend on the EOS and the microphysical effects taking place after the
merger. Interestingly, when considering the information collected with
the GW170817 event, some studies have constrained the maximum lifetime of
the remnant from that event to approximately one second \cite{Gill2019,
  Murguia-Berthier2020}. Overall, it is clear that the lifetime of the
HMNS can vary significantly between a few to several thousands of
milliseconds.
 
Here, we parametrize both decay times as power laws of the form
$\tau_\mrm{i} = \tau_\mrm{i,ref} ({\mathcal{M}_\mrm{chirp}}/
{\mathcal{M}_\mrm{chirp,ref}})^\gamma$ with $i = 1,\,2$, where
$\mchirp:=(m_1 m_2)^{3/5}/(m_1+m_2)^{1/5}$ is the chirp mass and $m_1$,
$m_2$ are the component masses. Note that $\gamma <0$, as higher chirp
masses correspond to a more rapid collapse, and reference decay times are
taken to be $\tau_{1} = 5\,{\rm ms}$ and $\tau_{2} = 50\,{\rm ms}$ which
correspond to typical lifetimes values Ref.~\cite{Rezzolla_main}. Because
a representative value of $\tau_{2}$ is only poorly known and the value
adopted above is quite conservative, we will consider later on how our
results are affected when considering larger values of
$\tau_{2}$. Finally, as a reference in our analysis we take the value
inferred for GW170817, i.e., $\mathcal{M}_\mrm{chirp,ref} =
1.186\,M_{\odot}$, and set $\gamma = -3$ (see below for a discussion of
the impact of varying $\tau_{2}$). The parameter $f_{1\epsilon} =
50\,{\rm Hz}$ introduces a small frequency modulation in the $f_1$ mode,
while $\beta_2$ adjusts the temporal phase of the waveform to ensure
smooth continuity between inspiral and postmerger phases. In this work we
set $\beta_2 = 0$ for simplicity and adjust the phase in the frequency
domain as explained below. Furthermore, higher-order contributions to the
waveform come from the coefficients $\gamma_2$ and $\xi_2$, but these
have a relatively minor impact on the total GW power. We take these
parameters to be constant and equal to their mean values over the four
EOSs studied by Ref.~\cite{Rezzolla_main}, thus $\gamma_2 = -554\,{\rm
  Hz}^2$ and $\xi_2 = 3.36 \times 10^4\,{\rm Hz}^3$.

\begin{figure}
    \centering
    \resizebox{\hsize}{!}{\includegraphics{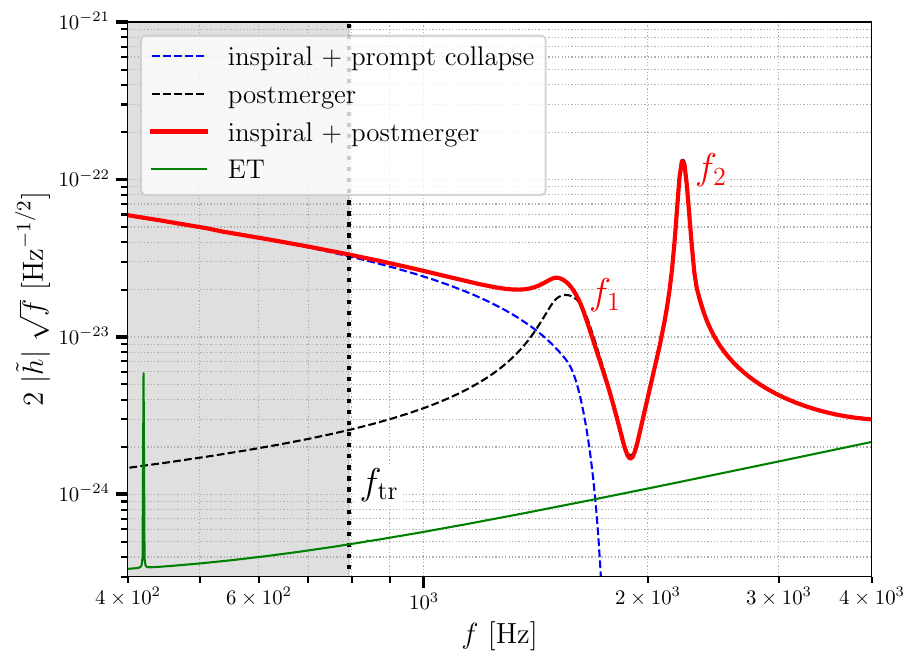}}
    \caption{Evolution in the frequency-domain of the GW strain for a BNS
      with masses $m_1=m_2=1.25\, M_{\odot}$ at a distance of observer $D
      = 50 \,{\rm Mpc}$ and relative to the GNH3
      EOS~\cite{Rezzolla_main}. The \texttt{IMRPhenomD\_NRTidalv2}
      waveform is shown with a black dashed line, the HMNS postmerger
      with a blue dashed line, and the combined waveform with a red solid
      line. The frequency $f_{\rm tr}$ marks the transition between the
      inspiral (grey-shaded are) and the postmerger, while the green
      solid curve reports the sensitivity of ET. Also reported are
        the main components $f_2$ and $f_1$ of the spectrum of equal-mass
        binaries.}
    \label{fig:WF_tot}
\end{figure}

Figure~\ref{fig:WF_tot} illustrates the construction of the hybrid
waveform for a BNS system with masses $M_1 = M_2 = 1.25\,M_{\odot}$, at a
distance from observer $D=50\,{\rm Mpc}$ when employing GNH3 EOS
discussed in Ref.~\cite{Rezzolla_main}. Shown with a blue dashed line is
the inspiral-and-merger signal followed by a prompt collapse and
described by the \texttt{IMRPhenomD\_NRTidalv2} model, the postmerger
signal with a black dashed line, and the combined (hybrid) signal with a
red solid line. We also show with a solid green line the ET strain
sensitivity. The transition frequency $f_{\rm tr}$ is meant to mark
therefore the boundary between the grey and white regions. In the grey
region, the hybrid waveform is simply given by the
\texttt{IMRPhenomD\_NRTidalv2} waveform. Note that in the white region,
the hybrid waveform is given by the sum of the inspiral and postmerger
strains and that the latter is negligible below $f_{\rm tr}$, thus
ensuring no significant power is lost or added. To properly combine the
two waveforms above $f_{\rm tr}$, we must ensure phase coherence, which
is not guaranteed a priori, as we have not explicitly computed the
temporal phase $\beta_2$. Therefore, at $f_{\rm tr}$, we adjust the phase
of the \texttt{IMRPhenomD\_NRTidalv2} waveform to match the postmerger
one, thus ensuring phase continuity.

{\it Modelling the BNS population}--- In order to calculate the SGWB from
the entire population of unresolved BNS mergers, we need to model the
merger rate as a function of redshift and system parameters. Here, we
consider (equal-mass) symmetric non-eccentric binaries, so that each BNS
is characterized by the component masses $m_1=m_2=m$, tidal deformability
parameters $\Lambda_1=\Lambda_2=\Lambda$, redshift $z$, sky position
(right ascension $\alpha$ and declination $\delta$), orbital inclination
angle $\iota$, polarization angle ($\psi$), coalescence phase
$\Phi_\mrm{c}$ and coalescence time set to be $t_\mrm{c}=0$. The mass
distribution we used is derived from Galactic observations and is assumed
to be independent of redshift. According to Ref.~\cite{farrow_mass_2019},
this distribution is well-modeled by a Gaussian with a mean of
$1.33\,M_\odot$ and a standard deviation of $0.09\,M_\odot$. Both neutron
stars in the binary are assumed to follow this distribution, which is
tightly concentrated around the mean, making it a non-critical parameter
for this study.

We note that this mass distribution may depend on the formation scenario
of the binary. More specifically, the mass distribution of BNS detected
via GW may differ from that of Galactic binaries, potentially explaining
the higher total mass observed in the BNS merger GW190425
\cite{2020ApJ...892L...3A, 2020MNRAS.496L..64R}. Furthermore, the recent
GW190814 event suggests a black hole-neutron star (BHNS) merger involving
a neutron star significantly heavier than typical Galactic counterparts
\cite{the_ligo_scientific_collaboration_gw190814_2020}, although there
are arguments favouring a black-hole origin for the
secondary~\cite{Nathanail2021}.

We assume that the BNS merger rate follows the star formation rate (SFR)
convolved with a distribution of delay times. Specifically, we adopt the
\texttt{baseline\_delays} model from~Ref.~\cite{LL2023}, where the merger
rate is defined as: $R_\mathrm{merg} (t) = R_0
\int^{t_{\mrm{d},\mathrm{max}}}_{t_{\mrm{d},\mathrm{min}}} \mrm{SFR}(t -
t_\mrm{d}) \, P(t_\mrm{d})\,\mrm{d}t_\mrm{d}$. Here the SFR is taken from
Ref.~\cite{vangioni_impact_2015} and $P(t_\mrm{d})$ is the distribution
of time intervals $t_\mrm{d}$ between the formation of stellar
progenitors and the eventual BNS merger. We here assume a power-law
distribution $P(t_\mrm{d})\propto t_\mrm{d}^{-1}$
\cite{2018MNRAS.474.2937C}, and the delay times to range from
$t_{\mrm{d},\mathrm{min}}=10\,{\rm Myr}$ to $t_{\mrm{d},\mathrm{max}}$
set by the Hubble time. The parameter $R_0$ is a normalization constant
that calibrates the merger rate at $z=0$ to the local BNS merger rate
inferred from the GWTC-3 LIGO-Virgo catalogue: $R_0=44^{+96}_{-34}~~{\rm
  yr}^{-1}\,{\rm
  Gpc}^{-3}$~\cite{the_ligo_scientific_collaboration_population_2022}.


\bibliographystyle{apsrev4-2}
\bibliography{biblio}

\begin{thebibliography}{105}%
\makeatletter
\providecommand \@ifxundefined [1]{%
 \@ifx{#1\undefined}
}%
\providecommand \@ifnum [1]{%
 \ifnum #1\expandafter \@firstoftwo
 \else \expandafter \@secondoftwo
 \fi
}%
\providecommand \@ifx [1]{%
 \ifx #1\expandafter \@firstoftwo
 \else \expandafter \@secondoftwo
 \fi
}%
\providecommand \natexlab [1]{#1}%
\providecommand \enquote  [1]{``#1''}%
\providecommand \bibnamefont  [1]{#1}%
\providecommand \bibfnamefont [1]{#1}%
\providecommand \citenamefont [1]{#1}%
\providecommand \href@noop [0]{\@secondoftwo}%
\providecommand \href [0]{\begingroup \@sanitize@url \@href}%
\providecommand \@href[1]{\@@startlink{#1}\@@href}%
\providecommand \@@href[1]{\endgroup#1\@@endlink}%
\providecommand \@sanitize@url [0]{\catcode `\\12\catcode `\$12\catcode
  `\&12\catcode `\#12\catcode `\^12\catcode `\_12\catcode `\%12\relax}%
\providecommand \@@startlink[1]{}%
\providecommand \@@endlink[0]{}%
\providecommand \url  [0]{\begingroup\@sanitize@url \@url }%
\providecommand \@url [1]{\endgroup\@href {#1}{\urlprefix }}%
\providecommand \urlprefix  [0]{URL }%
\providecommand \Eprint [0]{\href }%
\providecommand \doibase [0]{https://doi.org/}%
\providecommand \selectlanguage [0]{\@gobble}%
\providecommand \bibinfo  [0]{\@secondoftwo}%
\providecommand \bibfield  [0]{\@secondoftwo}%
\providecommand \translation [1]{[#1]}%
\providecommand \BibitemOpen [0]{}%
\providecommand \bibitemStop [0]{}%
\providecommand \bibitemNoStop [0]{.\EOS\space}%
\providecommand \EOS [0]{\spacefactor3000\relax}%
\providecommand \BibitemShut  [1]{\csname bibitem#1\endcsname}%
\let\auto@bib@innerbib\@empty
\bibitem [{\citenamefont {{LIGO and Virgo
  collaboration}}(2017{\natexlab{a}})}]{2017PhRvL.119p1101A}%
  \BibitemOpen
  \bibfield  {author} {\bibinfo {author} {\bibnamefont {{LIGO and Virgo
  collaboration}}},\ }\href {https://doi.org/10.1103/PhysRevLett.119.161101}
  {\bibfield  {journal} {\bibinfo  {journal} {Phys. Rev. Lett.}\ }\textbf
  {\bibinfo {volume} {119}},\ \bibinfo {eid} {161101} (\bibinfo {year}
  {2017}{\natexlab{a}})},\ \Eprint {https://arxiv.org/abs/1710.05832}
  {arXiv:1710.05832 [gr-qc]} \BibitemShut {NoStop}%
\bibitem [{\citenamefont {{LIGO collaboration}}(2015)}]{2015CQGra..32g4001L}%
  \BibitemOpen
  \bibfield  {author} {\bibinfo {author} {\bibnamefont {{LIGO
  collaboration}}},\ }\href {https://doi.org/10.1088/0264-9381/32/7/074001}
  {\bibfield  {journal} {\bibinfo  {journal} {Classical and Quantum Gravity}\
  }\textbf {\bibinfo {volume} {32}},\ \bibinfo {eid} {074001} (\bibinfo {year}
  {2015})},\ \Eprint {https://arxiv.org/abs/1411.4547} {arXiv:1411.4547
  [gr-qc]} \BibitemShut {NoStop}%
\bibitem [{\citenamefont {{Acernese, F. et al.}}(2015)}]{2015CQGra..32b4001A}%
  \BibitemOpen
  \bibfield  {author} {\bibinfo {author} {\bibnamefont {{Acernese, F. et
  al.}}},\ }\href {https://doi.org/10.1088/0264-9381/32/2/024001} {\bibfield
  {journal} {\bibinfo  {journal} {Classical and Quantum Gravity}\ }\textbf
  {\bibinfo {volume} {32}},\ \bibinfo {eid} {024001} (\bibinfo {year}
  {2015})},\ \Eprint {https://arxiv.org/abs/1408.3978} {arXiv:1408.3978
  [gr-qc]} \BibitemShut {NoStop}%
\bibitem [{\citenamefont {{Coulter, D.~A. et
  al.}}(2017)}]{2017Sci...358.1556C}%
  \BibitemOpen
  \bibfield  {author} {\bibinfo {author} {\bibnamefont {{Coulter, D.~A. et
  al.}}},\ }\href {https://doi.org/10.1126/science.aap9811} {\bibfield
  {journal} {\bibinfo  {journal} {Science}\ }\textbf {\bibinfo {volume}
  {358}},\ \bibinfo {pages} {1556} (\bibinfo {year} {2017})},\ \Eprint
  {https://arxiv.org/abs/1710.05452} {arXiv:1710.05452 [astro-ph.HE]}
  \BibitemShut {NoStop}%
\bibitem [{\citenamefont {{Goldstein, A. et al.}}(2017)}]{2017ApJ...848L..14G}%
  \BibitemOpen
  \bibfield  {author} {\bibinfo {author} {\bibnamefont {{Goldstein, A. et
  al.}}},\ }\href {https://doi.org/10.3847/2041-8213/aa8f41} {\bibfield
  {journal} {\bibinfo  {journal} {Astrophys. J. Lett.}\ }\textbf {\bibinfo
  {volume} {848}},\ \bibinfo {eid} {L14} (\bibinfo {year} {2017})},\ \Eprint
  {https://arxiv.org/abs/1710.05446} {arXiv:1710.05446 [astro-ph.HE]}
  \BibitemShut {NoStop}%
\bibitem [{\citenamefont {{Savchenko, V. et al.}}(2017)}]{2017ApJ...848L..15S}%
  \BibitemOpen
  \bibfield  {author} {\bibinfo {author} {\bibnamefont {{Savchenko, V. et
  al.}}},\ }\href {https://doi.org/10.3847/2041-8213/aa8f94} {\bibfield
  {journal} {\bibinfo  {journal} {Astrophys. J. Lett.}\ }\textbf {\bibinfo
  {volume} {848}},\ \bibinfo {eid} {L15} (\bibinfo {year} {2017})},\ \Eprint
  {https://arxiv.org/abs/1710.05449} {arXiv:1710.05449 [astro-ph.HE]}
  \BibitemShut {NoStop}%
\bibitem [{\citenamefont {{Mooley}}\ \emph {et~al.}(2018)\citenamefont
  {{Mooley}}, \citenamefont {{Deller}}, \citenamefont {{Gottlieb}},
  \citenamefont {{Nakar}}, \citenamefont {{Hallinan}}, \citenamefont
  {{Bourke}}, \citenamefont {{Frail}}, \citenamefont {{Horesh}}, \citenamefont
  {{Corsi}},\ and\ \citenamefont {{Hotokezaka}}}]{2018Natur.561..355M}%
  \BibitemOpen
  \bibfield  {author} {\bibinfo {author} {\bibfnamefont {K.~P.}\ \bibnamefont
  {{Mooley}}}, \bibinfo {author} {\bibfnamefont {A.~T.}\ \bibnamefont
  {{Deller}}}, \bibinfo {author} {\bibfnamefont {O.}~\bibnamefont
  {{Gottlieb}}}, \bibinfo {author} {\bibfnamefont {E.}~\bibnamefont {{Nakar}}},
  \bibinfo {author} {\bibfnamefont {G.}~\bibnamefont {{Hallinan}}}, \bibinfo
  {author} {\bibfnamefont {S.}~\bibnamefont {{Bourke}}}, \bibinfo {author}
  {\bibfnamefont {D.~A.}\ \bibnamefont {{Frail}}}, \bibinfo {author}
  {\bibfnamefont {A.}~\bibnamefont {{Horesh}}}, \bibinfo {author}
  {\bibfnamefont {A.}~\bibnamefont {{Corsi}}},\ and\ \bibinfo {author}
  {\bibfnamefont {K.}~\bibnamefont {{Hotokezaka}}},\ }\href
  {https://doi.org/10.1038/s41586-018-0486-3} {\bibfield  {journal} {\bibinfo
  {journal} {Nature}\ }\textbf {\bibinfo {volume} {561}},\ \bibinfo {pages}
  {355} (\bibinfo {year} {2018})},\ \Eprint {https://arxiv.org/abs/1806.09693}
  {arXiv:1806.09693 [astro-ph.HE]} \BibitemShut {NoStop}%
\bibitem [{\citenamefont {{Ghirlanda, G. et al.}}(2019)}]{2019Sci...363..968G}%
  \BibitemOpen
  \bibfield  {author} {\bibinfo {author} {\bibnamefont {{Ghirlanda, G. et
  al.}}},\ }\href {https://doi.org/10.1126/science.aau8815} {\bibfield
  {journal} {\bibinfo  {journal} {Science}\ }\textbf {\bibinfo {volume}
  {363}},\ \bibinfo {pages} {968} (\bibinfo {year} {2019})},\ \Eprint
  {https://arxiv.org/abs/1808.00469} {arXiv:1808.00469 [astro-ph.HE]}
  \BibitemShut {NoStop}%
\bibitem [{\citenamefont {collaboration}(2019)}]{2019PhRvL.123a1102A}%
  \BibitemOpen
  \bibfield  {author} {\bibinfo {author} {\bibfnamefont {L.}~\bibnamefont
  {collaboration}},\ }\href {https://doi.org/10.1103/PhysRevLett.123.011102}
  {\bibfield  {journal} {\bibinfo  {journal} {Phys. Rev. Lett.}\ }\textbf
  {\bibinfo {volume} {123}},\ \bibinfo {eid} {011102} (\bibinfo {year}
  {2019})},\ \Eprint {https://arxiv.org/abs/1811.00364} {arXiv:1811.00364
  [gr-qc]} \BibitemShut {NoStop}%
\bibitem [{\citenamefont {{LIGO and Virgo
  collaboration}}(2017{\natexlab{b}})}]{2017Natur.551...85A}%
  \BibitemOpen
  \bibfield  {author} {\bibinfo {author} {\bibnamefont {{LIGO and Virgo
  collaboration}}},\ }\href {https://doi.org/10.1038/nature24471} {\bibfield
  {journal} {\bibinfo  {journal} {Nature}\ }\textbf {\bibinfo {volume} {551}},\
  \bibinfo {pages} {85} (\bibinfo {year} {2017}{\natexlab{b}})},\ \Eprint
  {https://arxiv.org/abs/1710.05835} {arXiv:1710.05835 [astro-ph.CO]}
  \BibitemShut {NoStop}%
\bibitem [{\citenamefont {collaboration}(2020)}]{2020ApJ...892L...3A}%
  \BibitemOpen
  \bibfield  {author} {\bibinfo {author} {\bibfnamefont {L.}~\bibnamefont
  {collaboration}},\ }\href {https://doi.org/10.3847/2041-8213/ab75f5}
  {\bibfield  {journal} {\bibinfo  {journal} {Astrophys. J. Lett.}\ }\textbf
  {\bibinfo {volume} {892}},\ \bibinfo {eid} {L3} (\bibinfo {year} {2020})},\
  \Eprint {https://arxiv.org/abs/2001.01761} {arXiv:2001.01761 [astro-ph.HE]}
  \BibitemShut {NoStop}%
\bibitem [{\citenamefont {{Abbott, R. et
  al.}}(2023)}]{the_ligo_scientific_collaboration_population_2022}%
  \BibitemOpen
  \bibfield  {author} {\bibinfo {author} {\bibnamefont {{Abbott, R. et al.}}},\
  }\href {https://doi.org/10.1103/PhysRevX.13.011048} {\bibfield  {journal}
  {\bibinfo  {journal} {Physical Review X}\ }\textbf {\bibinfo {volume} {13}},\
  \bibinfo {eid} {011048} (\bibinfo {year} {2023})},\ \Eprint
  {https://arxiv.org/abs/2111.03634} {arXiv:2111.03634 [astro-ph.HE]}
  \BibitemShut {NoStop}%
\bibitem [{\citenamefont {{Bernuzzi}}\ \emph {et~al.}(2015)\citenamefont
  {{Bernuzzi}}, \citenamefont {{Nagar}}, \citenamefont {{Dietrich}},\ and\
  \citenamefont {{Damour}}}]{2015PhRvL.114p1103B}%
  \BibitemOpen
  \bibfield  {author} {\bibinfo {author} {\bibfnamefont {S.}~\bibnamefont
  {{Bernuzzi}}}, \bibinfo {author} {\bibfnamefont {A.}~\bibnamefont {{Nagar}}},
  \bibinfo {author} {\bibfnamefont {T.}~\bibnamefont {{Dietrich}}},\ and\
  \bibinfo {author} {\bibfnamefont {T.}~\bibnamefont {{Damour}}},\ }\href
  {https://doi.org/10.1103/PhysRevLett.114.161103} {\bibfield  {journal}
  {\bibinfo  {journal} {Phys. Rev. Lett.}\ }\textbf {\bibinfo {volume} {114}},\
  \bibinfo {eid} {161103} (\bibinfo {year} {2015})},\ \Eprint
  {https://arxiv.org/abs/1412.4553} {arXiv:1412.4553 [gr-qc]} \BibitemShut
  {NoStop}%
\bibitem [{\citenamefont {{Hinderer, T. et al.}}(2016)}]{2016PhRvL.116r1101H}%
  \BibitemOpen
  \bibfield  {author} {\bibinfo {author} {\bibnamefont {{Hinderer, T. et
  al.}}},\ }\href {https://doi.org/10.1103/PhysRevLett.116.181101} {\bibfield
  {journal} {\bibinfo  {journal} {Phys. Rev. Lett.}\ }\textbf {\bibinfo
  {volume} {116}},\ \bibinfo {eid} {181101} (\bibinfo {year} {2016})},\ \Eprint
  {https://arxiv.org/abs/1602.00599} {arXiv:1602.00599 [gr-qc]} \BibitemShut
  {NoStop}%
\bibitem [{\citenamefont {{Radice}}\ \emph {et~al.}(2014)\citenamefont
  {{Radice}}, \citenamefont {{Rezzolla}},\ and\ \citenamefont
  {{Galeazzi}}}]{2014MNRAS.437L..46R}%
  \BibitemOpen
  \bibfield  {author} {\bibinfo {author} {\bibfnamefont {D.}~\bibnamefont
  {{Radice}}}, \bibinfo {author} {\bibfnamefont {L.}~\bibnamefont
  {{Rezzolla}}},\ and\ \bibinfo {author} {\bibfnamefont {F.}~\bibnamefont
  {{Galeazzi}}},\ }\href {https://doi.org/10.1093/mnrasl/slt137} {\bibfield
  {journal} {\bibinfo  {journal} {Mon. Not. R. Astron. Soc.}\ }\textbf
  {\bibinfo {volume} {437}},\ \bibinfo {pages} {L46} (\bibinfo {year}
  {2014})},\ \Eprint {https://arxiv.org/abs/1306.6052} {arXiv:1306.6052
  [gr-qc]} \BibitemShut {NoStop}%
\bibitem [{\citenamefont {{Dietrich, Tim et al.}}(2019)}]{2019PhRvD..99b4029D}%
  \BibitemOpen
  \bibfield  {author} {\bibinfo {author} {\bibnamefont {{Dietrich, Tim et
  al.}}},\ }\href {https://doi.org/10.1103/PhysRevD.99.024029} {\bibfield
  {journal} {\bibinfo  {journal} {Phys. Rev. D}\ }\textbf {\bibinfo {volume}
  {99}},\ \bibinfo {eid} {024029} (\bibinfo {year} {2019})},\ \Eprint
  {https://arxiv.org/abs/1804.02235} {arXiv:1804.02235 [gr-qc]} \BibitemShut
  {NoStop}%
\bibitem [{\citenamefont {{Dietrich}}\ \emph {et~al.}(2019)\citenamefont
  {{Dietrich}}, \citenamefont {{Samajdar}}, \citenamefont {{Khan}},
  \citenamefont {{Johnson-McDaniel}}, \citenamefont {{Dudi}},\ and\
  \citenamefont {{Tichy}}}]{PhenomD_tidalv2}%
  \BibitemOpen
  \bibfield  {author} {\bibinfo {author} {\bibfnamefont {T.}~\bibnamefont
  {{Dietrich}}}, \bibinfo {author} {\bibfnamefont {A.}~\bibnamefont
  {{Samajdar}}}, \bibinfo {author} {\bibfnamefont {S.}~\bibnamefont {{Khan}}},
  \bibinfo {author} {\bibfnamefont {N.~K.}\ \bibnamefont {{Johnson-McDaniel}}},
  \bibinfo {author} {\bibfnamefont {R.}~\bibnamefont {{Dudi}}},\ and\ \bibinfo
  {author} {\bibfnamefont {W.}~\bibnamefont {{Tichy}}},\ }\href
  {https://doi.org/10.1103/PhysRevD.100.044003} {\bibfield  {journal} {\bibinfo
   {journal} {Phys. Rev. D}\ }\textbf {\bibinfo {volume} {100}},\ \bibinfo
  {eid} {044003} (\bibinfo {year} {2019})},\ \Eprint
  {https://arxiv.org/abs/1905.06011} {arXiv:1905.06011 [gr-qc]} \BibitemShut
  {NoStop}%
\bibitem [{\citenamefont {{Abbott, B.~P. et
  al.}}(2018{\natexlab{a}})}]{PhysRevLett.121.161101}%
  \BibitemOpen
  \bibfield  {author} {\bibinfo {author} {\bibnamefont {{Abbott, B.~P. et
  al.}}},\ }\href {https://doi.org/10.1103/PhysRevLett.121.161101} {\bibfield
  {journal} {\bibinfo  {journal} {Phys. Rev. Lett.}\ }\textbf {\bibinfo
  {volume} {121}},\ \bibinfo {eid} {161101} (\bibinfo {year}
  {2018}{\natexlab{a}})},\ \Eprint {https://arxiv.org/abs/1805.11581}
  {arXiv:1805.11581 [gr-qc]} \BibitemShut {NoStop}%
\bibitem [{\citenamefont {{LIGO and Virgo
  collaboration}}(2019)}]{2019PhRvX...9a1001A}%
  \BibitemOpen
  \bibfield  {author} {\bibinfo {author} {\bibnamefont {{LIGO and Virgo
  collaboration}}},\ }\href {https://doi.org/10.1103/PhysRevX.9.011001}
  {\bibfield  {journal} {\bibinfo  {journal} {Physical Review X}\ }\textbf
  {\bibinfo {volume} {9}},\ \bibinfo {eid} {011001} (\bibinfo {year} {2019})},\
  \Eprint {https://arxiv.org/abs/1805.11579} {arXiv:1805.11579 [gr-qc]}
  \BibitemShut {NoStop}%
\bibitem [{\citenamefont {{Shibata}}\ \emph {et~al.}(2017)\citenamefont
  {{Shibata}}, \citenamefont {{Fujibayashi}}, \citenamefont {{Hotokezaka}},
  \citenamefont {{Kiuchi}}, \citenamefont {{Kyutoku}}, \citenamefont
  {{Sekiguchi}},\ and\ \citenamefont {{Tanaka}}}]{2017PhRvD..96l3012S}%
  \BibitemOpen
  \bibfield  {author} {\bibinfo {author} {\bibfnamefont {M.}~\bibnamefont
  {{Shibata}}}, \bibinfo {author} {\bibfnamefont {S.}~\bibnamefont
  {{Fujibayashi}}}, \bibinfo {author} {\bibfnamefont {K.}~\bibnamefont
  {{Hotokezaka}}}, \bibinfo {author} {\bibfnamefont {K.}~\bibnamefont
  {{Kiuchi}}}, \bibinfo {author} {\bibfnamefont {K.}~\bibnamefont {{Kyutoku}}},
  \bibinfo {author} {\bibfnamefont {Y.}~\bibnamefont {{Sekiguchi}}},\ and\
  \bibinfo {author} {\bibfnamefont {M.}~\bibnamefont {{Tanaka}}},\ }\href
  {https://doi.org/10.1103/PhysRevD.96.123012} {\bibfield  {journal} {\bibinfo
  {journal} {Phys. Rev. D}\ }\textbf {\bibinfo {volume} {96}},\ \bibinfo {eid}
  {123012} (\bibinfo {year} {2017})},\ \Eprint
  {https://arxiv.org/abs/1710.07579} {arXiv:1710.07579 [astro-ph.HE]}
  \BibitemShut {NoStop}%
\bibitem [{\citenamefont {{Baiotti}}\ and\ \citenamefont
  {{Rezzolla}}(2017)}]{Rezzolla_bigReview}%
  \BibitemOpen
  \bibfield  {author} {\bibinfo {author} {\bibfnamefont {L.}~\bibnamefont
  {{Baiotti}}}\ and\ \bibinfo {author} {\bibfnamefont {L.}~\bibnamefont
  {{Rezzolla}}},\ }\href {https://doi.org/10.1088/1361-6633/aa67bb} {\bibfield
  {journal} {\bibinfo  {journal} {Reports on Progress in Physics}\ }\textbf
  {\bibinfo {volume} {80}},\ \bibinfo {eid} {096901} (\bibinfo {year}
  {2017})},\ \Eprint {https://arxiv.org/abs/1607.03540} {arXiv:1607.03540
  [gr-qc]} \BibitemShut {NoStop}%
\bibitem [{\citenamefont {{Radice}}\ \emph {et~al.}(2020)\citenamefont
  {{Radice}}, \citenamefont {{Bernuzzi}},\ and\ \citenamefont
  {{Perego}}}]{2020ARNPS..70...95R}%
  \BibitemOpen
  \bibfield  {author} {\bibinfo {author} {\bibfnamefont {D.}~\bibnamefont
  {{Radice}}}, \bibinfo {author} {\bibfnamefont {S.}~\bibnamefont
  {{Bernuzzi}}},\ and\ \bibinfo {author} {\bibfnamefont {A.}~\bibnamefont
  {{Perego}}},\ }\href {https://doi.org/10.1146/annurev-nucl-013120-114541}
  {\bibfield  {journal} {\bibinfo  {journal} {Annual Review of Nuclear and
  Particle Science}\ }\textbf {\bibinfo {volume} {70}},\ \bibinfo {pages} {95}
  (\bibinfo {year} {2020})},\ \Eprint {https://arxiv.org/abs/2002.03863}
  {arXiv:2002.03863 [astro-ph.HE]} \BibitemShut {NoStop}%
\bibitem [{\citenamefont {{Stergioulas}}\ \emph {et~al.}(2011)\citenamefont
  {{Stergioulas}}, \citenamefont {{Bauswein}}, \citenamefont {{Zagkouris}},\
  and\ \citenamefont {{Janka}}}]{2011MNRAS.418..427S}%
  \BibitemOpen
  \bibfield  {author} {\bibinfo {author} {\bibfnamefont {N.}~\bibnamefont
  {{Stergioulas}}}, \bibinfo {author} {\bibfnamefont {A.}~\bibnamefont
  {{Bauswein}}}, \bibinfo {author} {\bibfnamefont {K.}~\bibnamefont
  {{Zagkouris}}},\ and\ \bibinfo {author} {\bibfnamefont {H.-T.}\ \bibnamefont
  {{Janka}}},\ }\href {https://doi.org/10.1111/j.1365-2966.2011.19493.x}
  {\bibfield  {journal} {\bibinfo  {journal} {Mon. Not. R. Astron. Soc.}\
  }\textbf {\bibinfo {volume} {418}},\ \bibinfo {pages} {427} (\bibinfo {year}
  {2011})},\ \Eprint {https://arxiv.org/abs/1105.0368} {arXiv:1105.0368
  [gr-qc]} \BibitemShut {NoStop}%
\bibitem [{\citenamefont {{Bauswein}}\ and\ \citenamefont
  {{Janka}}(2012)}]{2012PhRvL.108a1101B}%
  \BibitemOpen
  \bibfield  {author} {\bibinfo {author} {\bibfnamefont {A.}~\bibnamefont
  {{Bauswein}}}\ and\ \bibinfo {author} {\bibfnamefont {H.~T.}\ \bibnamefont
  {{Janka}}},\ }\href {https://doi.org/10.1103/PhysRevLett.108.011101}
  {\bibfield  {journal} {\bibinfo  {journal} {Phys. Rev. Lett.}\ }\textbf
  {\bibinfo {volume} {108}},\ \bibinfo {eid} {011101} (\bibinfo {year}
  {2012})},\ \Eprint {https://arxiv.org/abs/1106.1616} {arXiv:1106.1616
  [astro-ph.SR]} \BibitemShut {NoStop}%
\bibitem [{\citenamefont {{Takami}}\ \emph
  {et~al.}(2014{\natexlab{a}})\citenamefont {{Takami}}, \citenamefont
  {{Rezzolla}},\ and\ \citenamefont {{Baiotti}}}]{2014PhRvL.113i1104T}%
  \BibitemOpen
  \bibfield  {author} {\bibinfo {author} {\bibfnamefont {K.}~\bibnamefont
  {{Takami}}}, \bibinfo {author} {\bibfnamefont {L.}~\bibnamefont
  {{Rezzolla}}},\ and\ \bibinfo {author} {\bibfnamefont {L.}~\bibnamefont
  {{Baiotti}}},\ }\href {https://doi.org/10.1103/PhysRevLett.113.091104}
  {\bibfield  {journal} {\bibinfo  {journal} {Phys. Rev. Lett.}\ }\textbf
  {\bibinfo {volume} {113}},\ \bibinfo {eid} {091104} (\bibinfo {year}
  {2014}{\natexlab{a}})},\ \Eprint {https://arxiv.org/abs/1403.5672}
  {arXiv:1403.5672 [gr-qc]} \BibitemShut {NoStop}%
\bibitem [{\citenamefont {{Bauswein}}\ and\ \citenamefont
  {{Stergioulas}}(2015{\natexlab{a}})}]{2015PhRvD..91l4056B}%
  \BibitemOpen
  \bibfield  {author} {\bibinfo {author} {\bibfnamefont {A.}~\bibnamefont
  {{Bauswein}}}\ and\ \bibinfo {author} {\bibfnamefont {N.}~\bibnamefont
  {{Stergioulas}}},\ }\href {https://doi.org/10.1103/PhysRevD.91.124056}
  {\bibfield  {journal} {\bibinfo  {journal} {Phys. Rev. D}\ }\textbf {\bibinfo
  {volume} {91}},\ \bibinfo {eid} {124056} (\bibinfo {year}
  {2015}{\natexlab{a}})},\ \Eprint {https://arxiv.org/abs/1502.03176}
  {arXiv:1502.03176 [astro-ph.SR]} \BibitemShut {NoStop}%
\bibitem [{\citenamefont {{Rezzolla}}\ and\ \citenamefont
  {{Takami}}(2016)}]{2016PhRvD..93l4051R}%
  \BibitemOpen
  \bibfield  {author} {\bibinfo {author} {\bibfnamefont {L.}~\bibnamefont
  {{Rezzolla}}}\ and\ \bibinfo {author} {\bibfnamefont {K.}~\bibnamefont
  {{Takami}}},\ }\href {https://doi.org/10.1103/PhysRevD.93.124051} {\bibfield
  {journal} {\bibinfo  {journal} {Phys. Rev. D}\ }\textbf {\bibinfo {volume}
  {93}},\ \bibinfo {eid} {124051} (\bibinfo {year} {2016})},\ \Eprint
  {https://arxiv.org/abs/1604.00246} {arXiv:1604.00246 [gr-qc]} \BibitemShut
  {NoStop}%
\bibitem [{\citenamefont {{Takami}}\ \emph
  {et~al.}(2015{\natexlab{a}})\citenamefont {{Takami}}, \citenamefont
  {{Rezzolla}},\ and\ \citenamefont {{Baiotti}}}]{2015PhRvD..91f4001T}%
  \BibitemOpen
  \bibfield  {author} {\bibinfo {author} {\bibfnamefont {K.}~\bibnamefont
  {{Takami}}}, \bibinfo {author} {\bibfnamefont {L.}~\bibnamefont
  {{Rezzolla}}},\ and\ \bibinfo {author} {\bibfnamefont {L.}~\bibnamefont
  {{Baiotti}}},\ }\href {https://doi.org/10.1103/PhysRevD.91.064001} {\bibfield
   {journal} {\bibinfo  {journal} {Phys. Rev. D}\ }\textbf {\bibinfo {volume}
  {91}},\ \bibinfo {eid} {064001} (\bibinfo {year} {2015}{\natexlab{a}})},\
  \Eprint {https://arxiv.org/abs/1412.3240} {arXiv:1412.3240 [gr-qc]}
  \BibitemShut {NoStop}%
\bibitem [{\citenamefont {{Topolski}}\ \emph {et~al.}(2024)\citenamefont
  {{Topolski}}, \citenamefont {{Tootle}},\ and\ \citenamefont
  {{Rezzolla}}}]{2024ApJ...960...86T}%
  \BibitemOpen
  \bibfield  {author} {\bibinfo {author} {\bibfnamefont {K.}~\bibnamefont
  {{Topolski}}}, \bibinfo {author} {\bibfnamefont {S.~D.}\ \bibnamefont
  {{Tootle}}},\ and\ \bibinfo {author} {\bibfnamefont {L.}~\bibnamefont
  {{Rezzolla}}},\ }\href {https://doi.org/10.3847/1538-4357/ad0152} {\bibfield
  {journal} {\bibinfo  {journal} {Astrophys. J.}\ }\textbf {\bibinfo {volume}
  {960}},\ \bibinfo {eid} {86} (\bibinfo {year} {2024})},\ \Eprint
  {https://arxiv.org/abs/2310.10728} {arXiv:2310.10728 [gr-qc]} \BibitemShut
  {NoStop}%
\bibitem [{\citenamefont {{Maggiore, M. et al.}}(2020)}]{2020JCAP...03..050M}%
  \BibitemOpen
  \bibfield  {author} {\bibinfo {author} {\bibnamefont {{Maggiore, M. et
  al.}}},\ }\href {https://doi.org/10.1088/1475-7516/2020/03/050} {\bibfield
  {journal} {\bibinfo  {journal} {Journal of Cosmology and Astroparticle
  Physics}\ }\textbf {\bibinfo {volume} {2020}}\bibfield  {number} {\bibinfo
  {number} { (3)},\ \bibinfo {eid} {050}},\ }\Eprint
  {https://arxiv.org/abs/1912.02622} {arXiv:1912.02622 [astro-ph.CO]}
  \BibitemShut {NoStop}%
\bibitem [{\citenamefont {{Evans, M. et al.}}(2021)}]{2021arXiv210909882E}%
  \BibitemOpen
  \bibfield  {author} {\bibinfo {author} {\bibnamefont {{Evans, M. et al.}}},\
  }\href {https://doi.org/10.48550/arXiv.2109.09882} {\bibfield  {journal}
  {\bibinfo  {journal} {arXiv e-prints}\ ,\ \bibinfo {eid} {arXiv:2109.09882}}
  (\bibinfo {year} {2021})},\ \Eprint {https://arxiv.org/abs/2109.09882}
  {arXiv:2109.09882 [astro-ph.IM]} \BibitemShut {NoStop}%
\bibitem [{\citenamefont {{Kiendrebeogo, R. et
  al.}}(2023)}]{2023ApJ...958..158K}%
  \BibitemOpen
  \bibfield  {author} {\bibinfo {author} {\bibnamefont {{Kiendrebeogo, R. et
  al.}}},\ }\href {https://doi.org/10.3847/1538-4357/acfcb1} {\bibfield
  {journal} {\bibinfo  {journal} {Astrophys. J.}\ }\textbf {\bibinfo {volume}
  {958}},\ \bibinfo {eid} {158} (\bibinfo {year} {2023})},\ \Eprint
  {https://arxiv.org/abs/2306.09234} {arXiv:2306.09234 [astro-ph.HE]}
  \BibitemShut {NoStop}%
\bibitem [{\citenamefont {{Branchesi M. et al.}}(2023)}]{2023JCAP...07..068B}%
  \BibitemOpen
  \bibfield  {author} {\bibinfo {author} {\bibnamefont {{Branchesi M. et
  al.}}},\ }\href {https://doi.org/10.1088/1475-7516/2023/07/068} {\bibfield
  {journal} {\bibinfo  {journal} {Journal of Cosmology and Astroparticle
  Physics}\ }\textbf {\bibinfo {volume} {2023}}\bibfield  {number} {\bibinfo
  {number} { (7)},\ \bibinfo {eid} {068}},\ }\Eprint
  {https://arxiv.org/abs/2303.15923} {arXiv:2303.15923 [gr-qc]} \BibitemShut
  {NoStop}%
\bibitem [{\citenamefont {{Breschi}}\ \emph {et~al.}(2022)\citenamefont
  {{Breschi}}, \citenamefont {{Gamba}}, \citenamefont {{Borhanian}},
  \citenamefont {{Carullo}},\ and\ \citenamefont
  {{Bernuzzi}}}]{2022arXiv220509979B}%
  \BibitemOpen
  \bibfield  {author} {\bibinfo {author} {\bibfnamefont {M.}~\bibnamefont
  {{Breschi}}}, \bibinfo {author} {\bibfnamefont {R.}~\bibnamefont {{Gamba}}},
  \bibinfo {author} {\bibfnamefont {S.}~\bibnamefont {{Borhanian}}}, \bibinfo
  {author} {\bibfnamefont {G.}~\bibnamefont {{Carullo}}},\ and\ \bibinfo
  {author} {\bibfnamefont {S.}~\bibnamefont {{Bernuzzi}}},\ }\href
  {https://doi.org/10.48550/arXiv.2205.09979} {\bibfield  {journal} {\bibinfo
  {journal} {arXiv e-prints}\ ,\ \bibinfo {eid} {arXiv:2205.09979}} (\bibinfo
  {year} {2022})},\ \Eprint {https://arxiv.org/abs/2205.09979}
  {arXiv:2205.09979 [gr-qc]} \BibitemShut {NoStop}%
\bibitem [{\citenamefont {{Regimbau}}(2011)}]{2011RAA....11..369R}%
  \BibitemOpen
  \bibfield  {author} {\bibinfo {author} {\bibfnamefont {T.}~\bibnamefont
  {{Regimbau}}},\ }\href {https://doi.org/10.1088/1674-4527/11/4/001}
  {\bibfield  {journal} {\bibinfo  {journal} {Research in Astronomy and
  Astrophysics}\ }\textbf {\bibinfo {volume} {11}},\ \bibinfo {pages} {369}
  (\bibinfo {year} {2011})},\ \Eprint {https://arxiv.org/abs/1101.2762}
  {arXiv:1101.2762 [astro-ph.CO]} \BibitemShut {NoStop}%
\bibitem [{\citenamefont {{Christensen}}(2019)}]{christensen_stochastic_2019}%
  \BibitemOpen
  \bibfield  {author} {\bibinfo {author} {\bibfnamefont {N.}~\bibnamefont
  {{Christensen}}},\ }\href {https://doi.org/10.1088/1361-6633/aae6b5}
  {\bibfield  {journal} {\bibinfo  {journal} {Reports on Progress in Physics}\
  }\textbf {\bibinfo {volume} {82}},\ \bibinfo {eid} {016903} (\bibinfo {year}
  {2019})},\ \Eprint {https://arxiv.org/abs/1811.08797} {arXiv:1811.08797
  [gr-qc]} \BibitemShut {NoStop}%
\bibitem [{\citenamefont {{Buonanno}}\ \emph {et~al.}(2005)\citenamefont
  {{Buonanno}}, \citenamefont {{Sigl}}, \citenamefont {{Raffelt}},
  \citenamefont {{Janka}},\ and\ \citenamefont
  {{M{\"u}ller}}}]{2005PhRvD..72h4001B}%
  \BibitemOpen
  \bibfield  {author} {\bibinfo {author} {\bibfnamefont {A.}~\bibnamefont
  {{Buonanno}}}, \bibinfo {author} {\bibfnamefont {G.}~\bibnamefont {{Sigl}}},
  \bibinfo {author} {\bibfnamefont {G.~G.}\ \bibnamefont {{Raffelt}}}, \bibinfo
  {author} {\bibfnamefont {H.-T.}\ \bibnamefont {{Janka}}},\ and\ \bibinfo
  {author} {\bibfnamefont {E.}~\bibnamefont {{M{\"u}ller}}},\ }\href
  {https://doi.org/10.1103/PhysRevD.72.084001} {\bibfield  {journal} {\bibinfo
  {journal} {Phys. Rev. D}\ }\textbf {\bibinfo {volume} {72}},\ \bibinfo {eid}
  {084001} (\bibinfo {year} {2005})},\ \Eprint
  {https://arxiv.org/abs/astro-ph/0412277} {arXiv:astro-ph/0412277 [astro-ph]}
  \BibitemShut {NoStop}%
\bibitem [{\citenamefont {{Regimbau}}\ and\ \citenamefont
  {{Mandic}}(2008)}]{regimbau_astrophysical_2008}%
  \BibitemOpen
  \bibfield  {author} {\bibinfo {author} {\bibfnamefont {T.}~\bibnamefont
  {{Regimbau}}}\ and\ \bibinfo {author} {\bibfnamefont {V.}~\bibnamefont
  {{Mandic}}},\ }\href {https://doi.org/10.1088/0264-9381/25/18/184018}
  {\bibfield  {journal} {\bibinfo  {journal} {Classical and Quantum Gravity}\
  }\textbf {\bibinfo {volume} {25}},\ \bibinfo {eid} {184018} (\bibinfo {year}
  {2008})},\ \Eprint {https://arxiv.org/abs/0806.2794} {arXiv:0806.2794
  [astro-ph]} \BibitemShut {NoStop}%
\bibitem [{\citenamefont {{Crocker}}\ \emph {et~al.}(2015)\citenamefont
  {{Crocker}}, \citenamefont {{Mandic}}, \citenamefont {{Regimbau}},
  \citenamefont {{Belczynski}}, \citenamefont {{Gladysz}}, \citenamefont
  {{Olive}}, \citenamefont {{Prestegard}},\ and\ \citenamefont
  {{Vangioni}}}]{2015PhRvD..92f3005C}%
  \BibitemOpen
  \bibfield  {author} {\bibinfo {author} {\bibfnamefont {K.}~\bibnamefont
  {{Crocker}}}, \bibinfo {author} {\bibfnamefont {V.}~\bibnamefont {{Mandic}}},
  \bibinfo {author} {\bibfnamefont {T.}~\bibnamefont {{Regimbau}}}, \bibinfo
  {author} {\bibfnamefont {K.}~\bibnamefont {{Belczynski}}}, \bibinfo {author}
  {\bibfnamefont {W.}~\bibnamefont {{Gladysz}}}, \bibinfo {author}
  {\bibfnamefont {K.}~\bibnamefont {{Olive}}}, \bibinfo {author} {\bibfnamefont
  {T.}~\bibnamefont {{Prestegard}}},\ and\ \bibinfo {author} {\bibfnamefont
  {E.}~\bibnamefont {{Vangioni}}},\ }\href
  {https://doi.org/10.1103/PhysRevD.92.063005} {\bibfield  {journal} {\bibinfo
  {journal} {Phys. Rev. D}\ }\textbf {\bibinfo {volume} {92}},\ \bibinfo {eid}
  {063005} (\bibinfo {year} {2015})},\ \Eprint
  {https://arxiv.org/abs/1506.02631} {arXiv:1506.02631 [gr-qc]} \BibitemShut
  {NoStop}%
\bibitem [{\citenamefont {{Regimbau}}(2022)}]{regimbau_quest_2022}%
  \BibitemOpen
  \bibfield  {author} {\bibinfo {author} {\bibfnamefont {T.}~\bibnamefont
  {{Regimbau}}},\ }\href {https://doi.org/10.3390/sym14020270} {\bibfield
  {journal} {\bibinfo  {journal} {Symmetry}\ }\textbf {\bibinfo {volume}
  {14}},\ \bibinfo {pages} {270} (\bibinfo {year} {2022})}\BibitemShut
  {NoStop}%
\bibitem [{\citenamefont {{Renzini}}\ \emph {et~al.}(2022)\citenamefont
  {{Renzini}}, \citenamefont {{Goncharov}}, \citenamefont {{Jenkins}},\ and\
  \citenamefont {{Meyers}}}]{2022Galax..10...34R}%
  \BibitemOpen
  \bibfield  {author} {\bibinfo {author} {\bibfnamefont {A.~I.}\ \bibnamefont
  {{Renzini}}}, \bibinfo {author} {\bibfnamefont {B.}~\bibnamefont
  {{Goncharov}}}, \bibinfo {author} {\bibfnamefont {A.~C.}\ \bibnamefont
  {{Jenkins}}},\ and\ \bibinfo {author} {\bibfnamefont {P.~M.}\ \bibnamefont
  {{Meyers}}},\ }\href {https://doi.org/10.3390/galaxies10010034} {\bibfield
  {journal} {\bibinfo  {journal} {Galaxies}\ }\textbf {\bibinfo {volume}
  {10}},\ \bibinfo {pages} {34} (\bibinfo {year} {2022})},\ \Eprint
  {https://arxiv.org/abs/2202.00178} {arXiv:2202.00178 [gr-qc]} \BibitemShut
  {NoStop}%
\bibitem [{\citenamefont {{Abbott, B.~P. et
  al.}}(2017)}]{PhysRevLett.119.029901}%
  \BibitemOpen
  \bibfield  {author} {\bibinfo {author} {\bibnamefont {{Abbott, B.~P. et
  al.}}},\ }\href {https://doi.org/10.1103/PhysRevLett.118.121101} {\bibfield
  {journal} {\bibinfo  {journal} {Phys. Rev. Lett.}\ }\textbf {\bibinfo
  {volume} {118}},\ \bibinfo {eid} {121101} (\bibinfo {year} {2017})},\ \Eprint
  {https://arxiv.org/abs/1612.02029} {arXiv:1612.02029 [gr-qc]} \BibitemShut
  {NoStop}%
\bibitem [{\citenamefont {{Abbott, B.~P. et
  al.}}(2018{\natexlab{b}})}]{PhysRevLett.120.091101}%
  \BibitemOpen
  \bibfield  {author} {\bibinfo {author} {\bibnamefont {{Abbott, B.~P. et
  al.}}},\ }\href {https://doi.org/10.1103/PhysRevLett.120.091101} {\bibfield
  {journal} {\bibinfo  {journal} {Phys. Rev. Lett.}\ }\textbf {\bibinfo
  {volume} {120}},\ \bibinfo {eid} {091101} (\bibinfo {year}
  {2018}{\natexlab{b}})},\ \Eprint {https://arxiv.org/abs/1710.05837}
  {arXiv:1710.05837 [gr-qc]} \BibitemShut {NoStop}%
\bibitem [{\citenamefont {{Abbott, B.~P. et al.}}(2019)}]{PhysRevD.100.061101}%
  \BibitemOpen
  \bibfield  {author} {\bibinfo {author} {\bibnamefont {{Abbott, B.~P. et
  al.}}},\ }\href {https://doi.org/10.1103/PhysRevD.100.061101} {\bibfield
  {journal} {\bibinfo  {journal} {Phys. Rev. D}\ }\textbf {\bibinfo {volume}
  {100}},\ \bibinfo {eid} {061101} (\bibinfo {year} {2019})},\ \Eprint
  {https://arxiv.org/abs/1903.02886} {arXiv:1903.02886 [gr-qc]} \BibitemShut
  {NoStop}%
\bibitem [{\citenamefont {{Abbott, B.~P. et al.}}(2021)}]{PhysRevD.104.022004}%
  \BibitemOpen
  \bibfield  {author} {\bibinfo {author} {\bibnamefont {{Abbott, B.~P. et
  al.}}},\ }\href {https://doi.org/10.1103/PhysRevD.104.022004} {\bibfield
  {journal} {\bibinfo  {journal} {Phys. Rev. D}\ }\textbf {\bibinfo {volume}
  {104}},\ \bibinfo {eid} {022004} (\bibinfo {year} {2021})},\ \Eprint
  {https://arxiv.org/abs/2101.12130} {arXiv:2101.12130 [gr-qc]} \BibitemShut
  {NoStop}%
\bibitem [{\citenamefont {{Abbott, R. et al.}}(2021)}]{PhysRevD.104.022005}%
  \BibitemOpen
  \bibfield  {author} {\bibinfo {author} {\bibnamefont {{Abbott, R. et al.}}},\
  }\href {https://doi.org/10.1103/PhysRevD.104.022005} {\bibfield  {journal}
  {\bibinfo  {journal} {Phys. Rev. D}\ }\textbf {\bibinfo {volume} {104}},\
  \bibinfo {eid} {022005} (\bibinfo {year} {2021})},\ \Eprint
  {https://arxiv.org/abs/2103.08520} {arXiv:2103.08520 [gr-qc]} \BibitemShut
  {NoStop}%
\bibitem [{\citenamefont {{Kowalska-Leszczynska}}\ \emph
  {et~al.}(2015)\citenamefont {{Kowalska-Leszczynska}}, \citenamefont
  {{Regimbau}}, \citenamefont {{Bulik}}, \citenamefont {{Dominik}},\ and\
  \citenamefont {{Belczynski}}}]{2015A&A...574A..58K}%
  \BibitemOpen
  \bibfield  {author} {\bibinfo {author} {\bibfnamefont {I.}~\bibnamefont
  {{Kowalska-Leszczynska}}}, \bibinfo {author} {\bibfnamefont {T.}~\bibnamefont
  {{Regimbau}}}, \bibinfo {author} {\bibfnamefont {T.}~\bibnamefont {{Bulik}}},
  \bibinfo {author} {\bibfnamefont {M.}~\bibnamefont {{Dominik}}},\ and\
  \bibinfo {author} {\bibfnamefont {K.}~\bibnamefont {{Belczynski}}},\ }\href
  {https://doi.org/10.1051/0004-6361/201424417} {\bibfield  {journal} {\bibinfo
   {journal} {Astron. Astrophys.}\ }\textbf {\bibinfo {volume} {574}},\
  \bibinfo {eid} {A58} (\bibinfo {year} {2015})},\ \Eprint
  {https://arxiv.org/abs/1205.4621} {arXiv:1205.4621 [astro-ph.CO]}
  \BibitemShut {NoStop}%
\bibitem [{\citenamefont {{Dvorkin}}\ \emph {et~al.}(2016)\citenamefont
  {{Dvorkin}}, \citenamefont {{Vangioni}}, \citenamefont {{Silk}},
  \citenamefont {{Uzan}},\ and\ \citenamefont
  {{Olive}}}]{dvorkin_metallicity-constrained_2016}%
  \BibitemOpen
  \bibfield  {author} {\bibinfo {author} {\bibfnamefont {I.}~\bibnamefont
  {{Dvorkin}}}, \bibinfo {author} {\bibfnamefont {E.}~\bibnamefont
  {{Vangioni}}}, \bibinfo {author} {\bibfnamefont {J.}~\bibnamefont {{Silk}}},
  \bibinfo {author} {\bibfnamefont {J.-P.}\ \bibnamefont {{Uzan}}},\ and\
  \bibinfo {author} {\bibfnamefont {K.~A.}\ \bibnamefont {{Olive}}},\ }\href
  {https://doi.org/10.1093/mnras/stw1477} {\bibfield  {journal} {\bibinfo
  {journal} {Mon. Not. R. Astron. Soc.}\ }\textbf {\bibinfo {volume} {461}},\
  \bibinfo {pages} {3877} (\bibinfo {year} {2016})},\ \Eprint
  {https://arxiv.org/abs/1604.04288} {arXiv:1604.04288 [astro-ph.HE]}
  \BibitemShut {NoStop}%
\bibitem [{\citenamefont {{Nakazato}}\ \emph {et~al.}(2016)\citenamefont
  {{Nakazato}}, \citenamefont {{Niino}},\ and\ \citenamefont
  {{Sago}}}]{Nakazato_2016}%
  \BibitemOpen
  \bibfield  {author} {\bibinfo {author} {\bibfnamefont {K.}~\bibnamefont
  {{Nakazato}}}, \bibinfo {author} {\bibfnamefont {Y.}~\bibnamefont
  {{Niino}}},\ and\ \bibinfo {author} {\bibfnamefont {N.}~\bibnamefont
  {{Sago}}},\ }\href {https://doi.org/10.3847/0004-637X/832/2/146} {\bibfield
  {journal} {\bibinfo  {journal} {Astrophys. J.}\ }\textbf {\bibinfo {volume}
  {832}},\ \bibinfo {eid} {146} (\bibinfo {year} {2016})},\ \Eprint
  {https://arxiv.org/abs/1605.02146} {arXiv:1605.02146 [astro-ph.HE]}
  \BibitemShut {NoStop}%
\bibitem [{\citenamefont {{P{\'e}rigois}}\ \emph {et~al.}(2021)\citenamefont
  {{P{\'e}rigois}}, \citenamefont {{Belczynski}}, \citenamefont {{Bulik}},\
  and\ \citenamefont {{Regimbau}}}]{perigois_startrack_2021}%
  \BibitemOpen
  \bibfield  {author} {\bibinfo {author} {\bibfnamefont {C.}~\bibnamefont
  {{P{\'e}rigois}}}, \bibinfo {author} {\bibfnamefont {C.}~\bibnamefont
  {{Belczynski}}}, \bibinfo {author} {\bibfnamefont {T.}~\bibnamefont
  {{Bulik}}},\ and\ \bibinfo {author} {\bibfnamefont {T.}~\bibnamefont
  {{Regimbau}}},\ }\href {https://doi.org/10.1103/PhysRevD.103.043002}
  {\bibfield  {journal} {\bibinfo  {journal} {Phys. Rev. D}\ }\textbf {\bibinfo
  {volume} {103}},\ \bibinfo {eid} {043002} (\bibinfo {year} {2021})},\ \Eprint
  {https://arxiv.org/abs/2008.04890} {arXiv:2008.04890 [astro-ph.CO]}
  \BibitemShut {NoStop}%
\bibitem [{\citenamefont {{Cusin}}\ \emph {et~al.}(2019)\citenamefont
  {{Cusin}}, \citenamefont {{Dvorkin}}, \citenamefont {{Pitrou}},\ and\
  \citenamefont {{Uzan}}}]{2019PhRvD.100f3004C}%
  \BibitemOpen
  \bibfield  {author} {\bibinfo {author} {\bibfnamefont {G.}~\bibnamefont
  {{Cusin}}}, \bibinfo {author} {\bibfnamefont {I.}~\bibnamefont {{Dvorkin}}},
  \bibinfo {author} {\bibfnamefont {C.}~\bibnamefont {{Pitrou}}},\ and\
  \bibinfo {author} {\bibfnamefont {J.-P.}\ \bibnamefont {{Uzan}}},\ }\href
  {https://doi.org/10.1103/PhysRevD.100.063004} {\bibfield  {journal} {\bibinfo
   {journal} {Phys. Rev. D}\ }\textbf {\bibinfo {volume} {100}},\ \bibinfo
  {eid} {063004} (\bibinfo {year} {2019})},\ \Eprint
  {https://arxiv.org/abs/1904.07797} {arXiv:1904.07797 [astro-ph.CO]}
  \BibitemShut {NoStop}%
\bibitem [{\citenamefont {{Callister}}\ \emph {et~al.}(2020)\citenamefont
  {{Callister}}, \citenamefont {{Fishbach}}, \citenamefont {{Holz}},\ and\
  \citenamefont {{Farr}}}]{2020ApJ...896L..32C}%
  \BibitemOpen
  \bibfield  {author} {\bibinfo {author} {\bibfnamefont {T.}~\bibnamefont
  {{Callister}}}, \bibinfo {author} {\bibfnamefont {M.}~\bibnamefont
  {{Fishbach}}}, \bibinfo {author} {\bibfnamefont {D.~E.}\ \bibnamefont
  {{Holz}}},\ and\ \bibinfo {author} {\bibfnamefont {W.~M.}\ \bibnamefont
  {{Farr}}},\ }\href {https://doi.org/10.3847/2041-8213/ab9743} {\bibfield
  {journal} {\bibinfo  {journal} {Astrophys. J. Lett.}\ }\textbf {\bibinfo
  {volume} {896}},\ \bibinfo {eid} {L32} (\bibinfo {year} {2020})},\ \Eprint
  {https://arxiv.org/abs/2003.12152} {arXiv:2003.12152 [astro-ph.HE]}
  \BibitemShut {NoStop}%
\bibitem [{\citenamefont {{P{\'e}rigois}}\ \emph {et~al.}(2022)\citenamefont
  {{P{\'e}rigois}}, \citenamefont {{Santoliquido}}, \citenamefont
  {{Bouffanais}}, \citenamefont {{Di Carlo}}, \citenamefont {{Giacobbo}},
  \citenamefont {{Rastello}}, \citenamefont {{Mapelli}},\ and\ \citenamefont
  {{Regimbau}}}]{perigois_gravitational_2022}%
  \BibitemOpen
  \bibfield  {author} {\bibinfo {author} {\bibfnamefont {C.}~\bibnamefont
  {{P{\'e}rigois}}}, \bibinfo {author} {\bibfnamefont {F.}~\bibnamefont
  {{Santoliquido}}}, \bibinfo {author} {\bibfnamefont {Y.}~\bibnamefont
  {{Bouffanais}}}, \bibinfo {author} {\bibfnamefont {U.~N.}\ \bibnamefont {{Di
  Carlo}}}, \bibinfo {author} {\bibfnamefont {N.}~\bibnamefont {{Giacobbo}}},
  \bibinfo {author} {\bibfnamefont {S.}~\bibnamefont {{Rastello}}}, \bibinfo
  {author} {\bibfnamefont {M.}~\bibnamefont {{Mapelli}}},\ and\ \bibinfo
  {author} {\bibfnamefont {T.}~\bibnamefont {{Regimbau}}},\ }\href
  {https://doi.org/10.1103/PhysRevD.105.103032} {\bibfield  {journal} {\bibinfo
   {journal} {Phys. Rev. D}\ }\textbf {\bibinfo {volume} {105}},\ \bibinfo
  {eid} {103032} (\bibinfo {year} {2022})},\ \Eprint
  {https://arxiv.org/abs/2112.01119} {arXiv:2112.01119 [astro-ph.CO]}
  \BibitemShut {NoStop}%
\bibitem [{\citenamefont {{Lehoucq}}\ \emph {et~al.}(2023)\citenamefont
  {{Lehoucq}}, \citenamefont {{Dvorkin}}, \citenamefont {{Srinivasan}},
  \citenamefont {{Pellouin}},\ and\ \citenamefont {{Lamberts}}}]{LL2023}%
  \BibitemOpen
  \bibfield  {author} {\bibinfo {author} {\bibfnamefont {L.}~\bibnamefont
  {{Lehoucq}}}, \bibinfo {author} {\bibfnamefont {I.}~\bibnamefont
  {{Dvorkin}}}, \bibinfo {author} {\bibfnamefont {R.}~\bibnamefont
  {{Srinivasan}}}, \bibinfo {author} {\bibfnamefont {C.}~\bibnamefont
  {{Pellouin}}},\ and\ \bibinfo {author} {\bibfnamefont {A.}~\bibnamefont
  {{Lamberts}}},\ }\href {https://doi.org/10.1093/mnras/stad2917} {\bibfield
  {journal} {\bibinfo  {journal} {Mon. Not. R. Astron. Soc.}\ }\textbf
  {\bibinfo {volume} {526}},\ \bibinfo {pages} {4378} (\bibinfo {year}
  {2023})},\ \Eprint {https://arxiv.org/abs/2306.09861} {arXiv:2306.09861
  [astro-ph.HE]} \BibitemShut {NoStop}%
\bibitem [{\citenamefont {{Annala}}\ \emph {et~al.}(2018)\citenamefont
  {{Annala}}, \citenamefont {{Gorda}}, \citenamefont {{Kurkela}},\ and\
  \citenamefont {{Vuorinen}}}]{Annala2017}%
  \BibitemOpen
  \bibfield  {author} {\bibinfo {author} {\bibfnamefont {E.}~\bibnamefont
  {{Annala}}}, \bibinfo {author} {\bibfnamefont {T.}~\bibnamefont {{Gorda}}},
  \bibinfo {author} {\bibfnamefont {A.}~\bibnamefont {{Kurkela}}},\ and\
  \bibinfo {author} {\bibfnamefont {A.}~\bibnamefont {{Vuorinen}}},\ }\href
  {https://doi.org/10.1103/PhysRevLett.120.172703} {\bibfield  {journal}
  {\bibinfo  {journal} {Phys. Rev. Lett.}\ }\textbf {\bibinfo {volume} {120}},\
  \bibinfo {eid} {172703} (\bibinfo {year} {2018})},\ \Eprint
  {https://arxiv.org/abs/1711.02644} {arXiv:1711.02644 [astro-ph.HE]}
  \BibitemShut {NoStop}%
\bibitem [{\citenamefont {{Most}}\ \emph {et~al.}(2018)\citenamefont {{Most}},
  \citenamefont {{Weih}}, \citenamefont {{Rezzolla}},\ and\ \citenamefont
  {{Schaffner-Bielich}}}]{Most2018}%
  \BibitemOpen
  \bibfield  {author} {\bibinfo {author} {\bibfnamefont {E.~R.}\ \bibnamefont
  {{Most}}}, \bibinfo {author} {\bibfnamefont {L.~R.}\ \bibnamefont {{Weih}}},
  \bibinfo {author} {\bibfnamefont {L.}~\bibnamefont {{Rezzolla}}},\ and\
  \bibinfo {author} {\bibfnamefont {J.}~\bibnamefont {{Schaffner-Bielich}}},\
  }\href {https://doi.org/10.1103/PhysRevLett.120.261103} {\bibfield  {journal}
  {\bibinfo  {journal} {Physical Review Letter}\ }\textbf {\bibinfo {volume}
  {120}},\ \bibinfo {eid} {261103} (\bibinfo {year} {2018})},\ \Eprint
  {https://arxiv.org/abs/1803.00549} {arXiv:1803.00549 [gr-qc]} \BibitemShut
  {NoStop}%
\bibitem [{\citenamefont {{Greif}}\ \emph {et~al.}(2019)\citenamefont
  {{Greif}}, \citenamefont {{Raaijmakers}}, \citenamefont {{Hebeler}},
  \citenamefont {{Schwenk}},\ and\ \citenamefont {{Watts}}}]{Greif2019}%
  \BibitemOpen
  \bibfield  {author} {\bibinfo {author} {\bibfnamefont {S.~K.}\ \bibnamefont
  {{Greif}}}, \bibinfo {author} {\bibfnamefont {G.}~\bibnamefont
  {{Raaijmakers}}}, \bibinfo {author} {\bibfnamefont {K.}~\bibnamefont
  {{Hebeler}}}, \bibinfo {author} {\bibfnamefont {A.}~\bibnamefont
  {{Schwenk}}},\ and\ \bibinfo {author} {\bibfnamefont {A.~L.}\ \bibnamefont
  {{Watts}}},\ }\href {https://doi.org/10.1093/mnras/stz654} {\bibfield
  {journal} {\bibinfo  {journal} {Monthly Notices of the Royal Astronomical
  Society}\ }\textbf {\bibinfo {volume} {485}},\ \bibinfo {pages} {5363}
  (\bibinfo {year} {2019})},\ \Eprint {https://arxiv.org/abs/1812.08188}
  {arXiv:1812.08188 [astro-ph.HE]} \BibitemShut {NoStop}%
\bibitem [{\citenamefont {{Altiparmak}}\ \emph {et~al.}(2022)\citenamefont
  {{Altiparmak}}, \citenamefont {{Ecker}},\ and\ \citenamefont
  {{Rezzolla}}}]{Rezzolla_EOS_1}%
  \BibitemOpen
  \bibfield  {author} {\bibinfo {author} {\bibfnamefont {S.}~\bibnamefont
  {{Altiparmak}}}, \bibinfo {author} {\bibfnamefont {C.}~\bibnamefont
  {{Ecker}}},\ and\ \bibinfo {author} {\bibfnamefont {L.}~\bibnamefont
  {{Rezzolla}}},\ }\href {https://doi.org/10.3847/2041-8213/ac9b2a} {\bibfield
  {journal} {\bibinfo  {journal} {Astrophys. J. Lett.}\ }\textbf {\bibinfo
  {volume} {939}},\ \bibinfo {eid} {L34} (\bibinfo {year} {2022})},\ \Eprint
  {https://arxiv.org/abs/2203.14974} {arXiv:2203.14974 [astro-ph.HE]}
  \BibitemShut {NoStop}%
\bibitem [{\citenamefont {{Ecker}}\ \emph {et~al.}(2025)\citenamefont
  {{Ecker}}, \citenamefont {{Gorda}}, \citenamefont {{Kurkela}},\ and\
  \citenamefont {{Rezzolla}}}]{ecker:2024b}%
  \BibitemOpen
  \bibfield  {author} {\bibinfo {author} {\bibfnamefont {C.}~\bibnamefont
  {{Ecker}}}, \bibinfo {author} {\bibfnamefont {T.}~\bibnamefont {{Gorda}}},
  \bibinfo {author} {\bibfnamefont {A.}~\bibnamefont {{Kurkela}}},\ and\
  \bibinfo {author} {\bibfnamefont {L.}~\bibnamefont {{Rezzolla}}},\ }\href
  {https://doi.org/10.1038/s41467-025-56500-x} {\bibfield  {journal} {\bibinfo
  {journal} {Nature Communications}\ }\textbf {\bibinfo {volume} {16}},\
  \bibinfo {eid} {1320} (\bibinfo {year} {2025})},\ \Eprint
  {https://arxiv.org/abs/2403.03246} {arXiv:2403.03246 [astro-ph.HE]}
  \BibitemShut {NoStop}%
\bibitem [{\citenamefont {{Bernuzzi}}\ \emph {et~al.}(2016)\citenamefont
  {{Bernuzzi}}, \citenamefont {{Radice}}, \citenamefont {{Ott}}, \citenamefont
  {{Roberts}}, \citenamefont {{Moesta}},\ and\ \citenamefont
  {{Galeazzi}}}]{bernuzzi2015b}%
  \BibitemOpen
  \bibfield  {author} {\bibinfo {author} {\bibfnamefont {S.}~\bibnamefont
  {{Bernuzzi}}}, \bibinfo {author} {\bibfnamefont {D.}~\bibnamefont
  {{Radice}}}, \bibinfo {author} {\bibfnamefont {C.~D.}\ \bibnamefont {{Ott}}},
  \bibinfo {author} {\bibfnamefont {L.~F.}\ \bibnamefont {{Roberts}}}, \bibinfo
  {author} {\bibfnamefont {P.}~\bibnamefont {{Moesta}}},\ and\ \bibinfo
  {author} {\bibfnamefont {F.}~\bibnamefont {{Galeazzi}}},\ }\href
  {https://doi.org/10.1103/PhysRevD.94.024023} {\bibfield  {journal} {\bibinfo
  {journal} {Phys. Rev. D}\ }\textbf {\bibinfo {volume} {94}},\ \bibinfo {eid}
  {024023} (\bibinfo {year} {2016})},\ \Eprint
  {https://arxiv.org/abs/1512.06397} {arXiv:1512.06397 [gr-qc]} \BibitemShut
  {NoStop}%
\bibitem [{\citenamefont {{Bose}}\ \emph {et~al.}(2018)\citenamefont {{Bose}},
  \citenamefont {{Chakravarti}}, \citenamefont {{Rezzolla}}, \citenamefont
  {{Sathyaprakash}},\ and\ \citenamefont {{Takami}}}]{Rezzolla_main}%
  \BibitemOpen
  \bibfield  {author} {\bibinfo {author} {\bibfnamefont {S.}~\bibnamefont
  {{Bose}}}, \bibinfo {author} {\bibfnamefont {K.}~\bibnamefont
  {{Chakravarti}}}, \bibinfo {author} {\bibfnamefont {L.}~\bibnamefont
  {{Rezzolla}}}, \bibinfo {author} {\bibfnamefont {B.~S.}\ \bibnamefont
  {{Sathyaprakash}}},\ and\ \bibinfo {author} {\bibfnamefont {K.}~\bibnamefont
  {{Takami}}},\ }\href {https://doi.org/10.1103/PhysRevLett.120.031102}
  {\bibfield  {journal} {\bibinfo  {journal} {Phys. Rev. Lett.}\ }\textbf
  {\bibinfo {volume} {120}},\ \bibinfo {eid} {031102} (\bibinfo {year}
  {2018})},\ \Eprint {https://arxiv.org/abs/1705.10850} {arXiv:1705.10850
  [gr-qc]} \BibitemShut {NoStop}%
\bibitem [{\citenamefont {{Ecker}}\ and\ \citenamefont
  {{Rezzolla}}(2022)}]{Rezzolla_EOS_2}%
  \BibitemOpen
  \bibfield  {author} {\bibinfo {author} {\bibfnamefont {C.}~\bibnamefont
  {{Ecker}}}\ and\ \bibinfo {author} {\bibfnamefont {L.}~\bibnamefont
  {{Rezzolla}}},\ }\href {https://doi.org/10.3847/2041-8213/ac8674} {\bibfield
  {journal} {\bibinfo  {journal} {Astrophys. J. Lett.}\ }\textbf {\bibinfo
  {volume} {939}},\ \bibinfo {eid} {L35} (\bibinfo {year} {2022})},\ \Eprint
  {https://arxiv.org/abs/2207.04417} {arXiv:2207.04417 [gr-qc]} \BibitemShut
  {NoStop}%
\bibitem [{\citenamefont {{Ecker}}\ and\ \citenamefont
  {{Rezzolla}}(2023)}]{Rezzolla_EOS_3}%
  \BibitemOpen
  \bibfield  {author} {\bibinfo {author} {\bibfnamefont {C.}~\bibnamefont
  {{Ecker}}}\ and\ \bibinfo {author} {\bibfnamefont {L.}~\bibnamefont
  {{Rezzolla}}},\ }\href {https://doi.org/10.1093/mnras/stac3755} {\bibfield
  {journal} {\bibinfo  {journal} {Mon. Not. R. Astron. Soc.}\ }\textbf
  {\bibinfo {volume} {519}},\ \bibinfo {pages} {2615} (\bibinfo {year}
  {2023})},\ \Eprint {https://arxiv.org/abs/2209.08101} {arXiv:2209.08101
  [astro-ph.HE]} \BibitemShut {NoStop}%
\bibitem [{\citenamefont {Jiang}\ \emph {et~al.}(2023)\citenamefont {Jiang},
  \citenamefont {Ecker},\ and\ \citenamefont {Rezzolla}}]{Rezzolla_EOS_4}%
  \BibitemOpen
  \bibfield  {author} {\bibinfo {author} {\bibfnamefont {J.-L.}\ \bibnamefont
  {Jiang}}, \bibinfo {author} {\bibfnamefont {C.}~\bibnamefont {Ecker}},\ and\
  \bibinfo {author} {\bibfnamefont {L.}~\bibnamefont {Rezzolla}},\ }\href
  {https://doi.org/10.3847/1538-4357/acc4be} {\bibfield  {journal} {\bibinfo
  {journal} {The Astrophysical Journal}\ }\textbf {\bibinfo {volume} {949}},\
  \bibinfo {pages} {11} (\bibinfo {year} {2023})}\BibitemShut {NoStop}%
\bibitem [{\citenamefont {{Allen}}\ and\ \citenamefont
  {{Romano}}(1999)}]{1999PhRvD..59j2001A}%
  \BibitemOpen
  \bibfield  {author} {\bibinfo {author} {\bibfnamefont {B.}~\bibnamefont
  {{Allen}}}\ and\ \bibinfo {author} {\bibfnamefont {J.~D.}\ \bibnamefont
  {{Romano}}},\ }\href {https://doi.org/10.1103/PhysRevD.59.102001} {\bibfield
  {journal} {\bibinfo  {journal} {Phys. Rev. D}\ }\textbf {\bibinfo {volume}
  {59}},\ \bibinfo {eid} {102001} (\bibinfo {year} {1999})},\ \Eprint
  {https://arxiv.org/abs/gr-qc/9710117} {arXiv:gr-qc/9710117 [gr-qc]}
  \BibitemShut {NoStop}%
\bibitem [{\citenamefont {{Margalit}}\ and\ \citenamefont
  {{Metzger}}(2017)}]{Margalit2017}%
  \BibitemOpen
  \bibfield  {author} {\bibinfo {author} {\bibfnamefont {B.}~\bibnamefont
  {{Margalit}}}\ and\ \bibinfo {author} {\bibfnamefont {B.~D.}\ \bibnamefont
  {{Metzger}}},\ }\href {https://doi.org/10.3847/2041-8213/aa991c} {\bibfield
  {journal} {\bibinfo  {journal} {Astrophys. J. Lett.}\ }\textbf {\bibinfo
  {volume} {850}},\ \bibinfo {eid} {L19} (\bibinfo {year} {2017})},\ \Eprint
  {https://arxiv.org/abs/1710.05938} {arXiv:1710.05938 [astro-ph.HE]}
  \BibitemShut {NoStop}%
\bibitem [{\citenamefont {{Rezzolla}}\ \emph {et~al.}(2018)\citenamefont
  {{Rezzolla}}, \citenamefont {{Most}},\ and\ \citenamefont
  {{Weih}}}]{Rezzolla2017}%
  \BibitemOpen
  \bibfield  {author} {\bibinfo {author} {\bibfnamefont {L.}~\bibnamefont
  {{Rezzolla}}}, \bibinfo {author} {\bibfnamefont {E.~R.}\ \bibnamefont
  {{Most}}},\ and\ \bibinfo {author} {\bibfnamefont {L.~R.}\ \bibnamefont
  {{Weih}}},\ }\href {https://doi.org/10.3847/2041-8213/aaa401} {\bibfield
  {journal} {\bibinfo  {journal} {Astrophys. J. Lett.}\ }\textbf {\bibinfo
  {volume} {852}},\ \bibinfo {eid} {L25} (\bibinfo {year} {2018})},\ \Eprint
  {https://arxiv.org/abs/1711.00314} {arXiv:1711.00314 [astro-ph.HE]}
  \BibitemShut {NoStop}%
\bibitem [{\citenamefont {{Ruiz}}\ \emph {et~al.}(2018)\citenamefont {{Ruiz}},
  \citenamefont {{Shapiro}},\ and\ \citenamefont {{Tsokaros}}}]{Ruiz2017}%
  \BibitemOpen
  \bibfield  {author} {\bibinfo {author} {\bibfnamefont {M.}~\bibnamefont
  {{Ruiz}}}, \bibinfo {author} {\bibfnamefont {S.~L.}\ \bibnamefont
  {{Shapiro}}},\ and\ \bibinfo {author} {\bibfnamefont {A.}~\bibnamefont
  {{Tsokaros}}},\ }\href {https://doi.org/10.1103/PhysRevD.97.021501}
  {\bibfield  {journal} {\bibinfo  {journal} {Phys. Rev. D}\ }\textbf {\bibinfo
  {volume} {97}},\ \bibinfo {eid} {021501} (\bibinfo {year} {2018})},\ \Eprint
  {https://arxiv.org/abs/1711.00473} {arXiv:1711.00473 [astro-ph.HE]}
  \BibitemShut {NoStop}%
\bibitem [{\citenamefont {{Shibata}}\ \emph {et~al.}(2019)\citenamefont
  {{Shibata}}, \citenamefont {{Zhou}}, \citenamefont {{Kiuchi}},\ and\
  \citenamefont {{Fujibayashi}}}]{Shibata2019}%
  \BibitemOpen
  \bibfield  {author} {\bibinfo {author} {\bibfnamefont {M.}~\bibnamefont
  {{Shibata}}}, \bibinfo {author} {\bibfnamefont {E.}~\bibnamefont {{Zhou}}},
  \bibinfo {author} {\bibfnamefont {K.}~\bibnamefont {{Kiuchi}}},\ and\
  \bibinfo {author} {\bibfnamefont {S.}~\bibnamefont {{Fujibayashi}}},\ }\href
  {https://doi.org/10.1103/PhysRevD.100.023015} {\bibfield  {journal} {\bibinfo
   {journal} {Phys. Rev. D}\ }\textbf {\bibinfo {volume} {100}},\ \bibinfo
  {eid} {023015} (\bibinfo {year} {2019})},\ \Eprint
  {https://arxiv.org/abs/1905.03656} {arXiv:1905.03656 [astro-ph.HE]}
  \BibitemShut {NoStop}%
\bibitem [{\citenamefont {{Nathanail}}\ \emph {et~al.}(2021)\citenamefont
  {{Nathanail}}, \citenamefont {{Most}},\ and\ \citenamefont
  {{Rezzolla}}}]{Nathanail2021}%
  \BibitemOpen
  \bibfield  {author} {\bibinfo {author} {\bibfnamefont {A.}~\bibnamefont
  {{Nathanail}}}, \bibinfo {author} {\bibfnamefont {E.~R.}\ \bibnamefont
  {{Most}}},\ and\ \bibinfo {author} {\bibfnamefont {L.}~\bibnamefont
  {{Rezzolla}}},\ }\href {https://doi.org/10.3847/2041-8213/abdfc6} {\bibfield
  {journal} {\bibinfo  {journal} {Astrophys. J. Lett.}\ }\textbf {\bibinfo
  {volume} {908}},\ \bibinfo {eid} {L28} (\bibinfo {year} {2021})},\ \Eprint
  {https://arxiv.org/abs/2101.01735} {arXiv:2101.01735 [astro-ph.HE]}
  \BibitemShut {NoStop}%
\bibitem [{\citenamefont {{Vangioni}}\ \emph {et~al.}(2015)\citenamefont
  {{Vangioni}}, \citenamefont {{Olive}}, \citenamefont {{Prestegard}},
  \citenamefont {{Silk}}, \citenamefont {{Petitjean}},\ and\ \citenamefont
  {{Mandic}}}]{vangioni_impact_2015}%
  \BibitemOpen
  \bibfield  {author} {\bibinfo {author} {\bibfnamefont {E.}~\bibnamefont
  {{Vangioni}}}, \bibinfo {author} {\bibfnamefont {K.~A.}\ \bibnamefont
  {{Olive}}}, \bibinfo {author} {\bibfnamefont {T.}~\bibnamefont
  {{Prestegard}}}, \bibinfo {author} {\bibfnamefont {J.}~\bibnamefont
  {{Silk}}}, \bibinfo {author} {\bibfnamefont {P.}~\bibnamefont
  {{Petitjean}}},\ and\ \bibinfo {author} {\bibfnamefont {V.}~\bibnamefont
  {{Mandic}}},\ }\href {https://doi.org/10.1093/mnras/stu2600} {\bibfield
  {journal} {\bibinfo  {journal} {Mon. Not. R. Astron. Soc.}\ }\textbf
  {\bibinfo {volume} {447}},\ \bibinfo {pages} {2575} (\bibinfo {year}
  {2015})},\ \Eprint {https://arxiv.org/abs/1409.2462} {arXiv:1409.2462
  [astro-ph.GA]} \BibitemShut {NoStop}%
\bibitem [{\citenamefont {{Chruslinska}}\ \emph {et~al.}(2018)\citenamefont
  {{Chruslinska}}, \citenamefont {{Belczynski}}, \citenamefont {{Klencki}},\
  and\ \citenamefont {{Benacquista}}}]{2018MNRAS.474.2937C}%
  \BibitemOpen
  \bibfield  {author} {\bibinfo {author} {\bibfnamefont {M.}~\bibnamefont
  {{Chruslinska}}}, \bibinfo {author} {\bibfnamefont {K.}~\bibnamefont
  {{Belczynski}}}, \bibinfo {author} {\bibfnamefont {J.}~\bibnamefont
  {{Klencki}}},\ and\ \bibinfo {author} {\bibfnamefont {M.}~\bibnamefont
  {{Benacquista}}},\ }\href {https://doi.org/10.1093/mnras/stx2923} {\bibfield
  {journal} {\bibinfo  {journal} {Mon. Not. R. Astron. Soc.}\ }\textbf
  {\bibinfo {volume} {474}},\ \bibinfo {pages} {2937} (\bibinfo {year}
  {2018})},\ \Eprint {https://arxiv.org/abs/1708.07885} {arXiv:1708.07885
  [astro-ph.HE]} \BibitemShut {NoStop}%
\bibitem [{\citenamefont {{Iacovelli}}\ \emph {et~al.}(2022)\citenamefont
  {{Iacovelli}}, \citenamefont {{Mancarella}}, \citenamefont {{Foffa}},\ and\
  \citenamefont {{Maggiore}}}]{2022ApJ...941..208I}%
  \BibitemOpen
  \bibfield  {author} {\bibinfo {author} {\bibfnamefont {F.}~\bibnamefont
  {{Iacovelli}}}, \bibinfo {author} {\bibfnamefont {M.}~\bibnamefont
  {{Mancarella}}}, \bibinfo {author} {\bibfnamefont {S.}~\bibnamefont
  {{Foffa}}},\ and\ \bibinfo {author} {\bibfnamefont {M.}~\bibnamefont
  {{Maggiore}}},\ }\href {https://doi.org/10.3847/1538-4357/ac9cd4} {\bibfield
  {journal} {\bibinfo  {journal} {Astrophys. J.}\ }\textbf {\bibinfo {volume}
  {941}},\ \bibinfo {eid} {208} (\bibinfo {year} {2022})},\ \Eprint
  {https://arxiv.org/abs/2207.02771} {arXiv:2207.02771 [gr-qc]} \BibitemShut
  {NoStop}%
\bibitem [{\citenamefont {{Dupletsa}}\ \emph {et~al.}(2023)\citenamefont
  {{Dupletsa}}, \citenamefont {{Harms}}, \citenamefont {{Banerjee}},
  \citenamefont {{Branchesi}}, \citenamefont {{Goncharov}}, \citenamefont
  {{Maselli}}, \citenamefont {{Oliveira}}, \citenamefont {{Ronchini}},\ and\
  \citenamefont {{Tissino}}}]{GWFish}%
  \BibitemOpen
  \bibfield  {author} {\bibinfo {author} {\bibfnamefont {U.}~\bibnamefont
  {{Dupletsa}}}, \bibinfo {author} {\bibfnamefont {J.}~\bibnamefont {{Harms}}},
  \bibinfo {author} {\bibfnamefont {B.}~\bibnamefont {{Banerjee}}}, \bibinfo
  {author} {\bibfnamefont {M.}~\bibnamefont {{Branchesi}}}, \bibinfo {author}
  {\bibfnamefont {B.}~\bibnamefont {{Goncharov}}}, \bibinfo {author}
  {\bibfnamefont {A.}~\bibnamefont {{Maselli}}}, \bibinfo {author}
  {\bibfnamefont {A.~C.~S.}\ \bibnamefont {{Oliveira}}}, \bibinfo {author}
  {\bibfnamefont {S.}~\bibnamefont {{Ronchini}}},\ and\ \bibinfo {author}
  {\bibfnamefont {J.}~\bibnamefont {{Tissino}}},\ }\href
  {https://doi.org/10.1016/j.ascom.2022.100671} {\bibfield  {journal} {\bibinfo
   {journal} {Astronomy and Computing}\ }\textbf {\bibinfo {volume} {42}},\
  \bibinfo {eid} {100671} (\bibinfo {year} {2023})},\ \Eprint
  {https://arxiv.org/abs/2205.02499} {arXiv:2205.02499 [gr-qc]} \BibitemShut
  {NoStop}%
\bibitem [{\citenamefont {{Sachdev}}\ \emph
  {et~al.}(2020{\natexlab{a}})\citenamefont {{Sachdev}}, \citenamefont
  {{Regimbau}},\ and\ \citenamefont
  {{Sathyaprakash}}}]{background_subtraction}%
  \BibitemOpen
  \bibfield  {author} {\bibinfo {author} {\bibfnamefont {S.}~\bibnamefont
  {{Sachdev}}}, \bibinfo {author} {\bibfnamefont {T.}~\bibnamefont
  {{Regimbau}}},\ and\ \bibinfo {author} {\bibfnamefont {B.~S.}\ \bibnamefont
  {{Sathyaprakash}}},\ }\href {https://doi.org/10.1103/PhysRevD.102.024051}
  {\bibfield  {journal} {\bibinfo  {journal} {Phys. Rev. D}\ }\textbf {\bibinfo
  {volume} {102}},\ \bibinfo {eid} {024051} (\bibinfo {year}
  {2020}{\natexlab{a}})},\ \Eprint {https://arxiv.org/abs/2002.05365}
  {arXiv:2002.05365 [gr-qc]} \BibitemShut {NoStop}%
\bibitem [{\citenamefont {{Regimbau}}\ \emph {et~al.}(2017)\citenamefont
  {{Regimbau}}, \citenamefont {{Evans}}, \citenamefont {{Christensen}},
  \citenamefont {{Katsavounidis}}, \citenamefont {{Sathyaprakash}},\ and\
  \citenamefont {{Vitale}}}]{2017PhRvL.118o1105R}%
  \BibitemOpen
  \bibfield  {author} {\bibinfo {author} {\bibfnamefont {T.}~\bibnamefont
  {{Regimbau}}}, \bibinfo {author} {\bibfnamefont {M.}~\bibnamefont {{Evans}}},
  \bibinfo {author} {\bibfnamefont {N.}~\bibnamefont {{Christensen}}}, \bibinfo
  {author} {\bibfnamefont {E.}~\bibnamefont {{Katsavounidis}}}, \bibinfo
  {author} {\bibfnamefont {B.}~\bibnamefont {{Sathyaprakash}}},\ and\ \bibinfo
  {author} {\bibfnamefont {S.}~\bibnamefont {{Vitale}}},\ }\href
  {https://doi.org/10.1103/PhysRevLett.118.151105} {\bibfield  {journal}
  {\bibinfo  {journal} {Phys. Rev. Lett.}\ }\textbf {\bibinfo {volume} {118}},\
  \bibinfo {eid} {151105} (\bibinfo {year} {2017})},\ \Eprint
  {https://arxiv.org/abs/1611.08943} {arXiv:1611.08943 [astro-ph.CO]}
  \BibitemShut {NoStop}%
\bibitem [{\citenamefont {{Sachdev}}\ \emph
  {et~al.}(2020{\natexlab{b}})\citenamefont {{Sachdev}}, \citenamefont
  {{Regimbau}},\ and\ \citenamefont {{Sathyaprakash}}}]{2020PhRvD.102b4051S}%
  \BibitemOpen
  \bibfield  {author} {\bibinfo {author} {\bibfnamefont {S.}~\bibnamefont
  {{Sachdev}}}, \bibinfo {author} {\bibfnamefont {T.}~\bibnamefont
  {{Regimbau}}},\ and\ \bibinfo {author} {\bibfnamefont {B.~S.}\ \bibnamefont
  {{Sathyaprakash}}},\ }\href {https://doi.org/10.1103/PhysRevD.102.024051}
  {\bibfield  {journal} {\bibinfo  {journal} {Phys. Rev. D}\ }\textbf {\bibinfo
  {volume} {102}},\ \bibinfo {eid} {024051} (\bibinfo {year}
  {2020}{\natexlab{b}})},\ \Eprint {https://arxiv.org/abs/2002.05365}
  {arXiv:2002.05365 [gr-qc]} \BibitemShut {NoStop}%
\bibitem [{\citenamefont {{Zhou}}\ \emph {et~al.}(2023)\citenamefont {{Zhou}},
  \citenamefont {{Reali}}, \citenamefont {{Berti}}, \citenamefont
  {{{\c{c}}al{\i}{\c{s}}kan}}, \citenamefont {{Creque-Sarbinowski}},
  \citenamefont {{Kamionkowski}},\ and\ \citenamefont
  {{Sathyaprakash}}}]{Pop+omegaerr}%
  \BibitemOpen
  \bibfield  {author} {\bibinfo {author} {\bibfnamefont {B.}~\bibnamefont
  {{Zhou}}}, \bibinfo {author} {\bibfnamefont {L.}~\bibnamefont {{Reali}}},
  \bibinfo {author} {\bibfnamefont {E.}~\bibnamefont {{Berti}}}, \bibinfo
  {author} {\bibfnamefont {M.}~\bibnamefont {{{\c{c}}al{\i}{\c{s}}kan}}},
  \bibinfo {author} {\bibfnamefont {C.}~\bibnamefont {{Creque-Sarbinowski}}},
  \bibinfo {author} {\bibfnamefont {M.}~\bibnamefont {{Kamionkowski}}},\ and\
  \bibinfo {author} {\bibfnamefont {B.~S.}\ \bibnamefont {{Sathyaprakash}}},\
  }\href {https://doi.org/10.1103/PhysRevD.108.064040} {\bibfield  {journal}
  {\bibinfo  {journal} {Phys. Rev. D}\ }\textbf {\bibinfo {volume} {108}},\
  \bibinfo {eid} {064040} (\bibinfo {year} {2023})},\ \Eprint
  {https://arxiv.org/abs/2209.01310} {arXiv:2209.01310 [gr-qc]} \BibitemShut
  {NoStop}%
\bibitem [{\citenamefont {{Meacher}}\ \emph {et~al.}(2015)\citenamefont
  {{Meacher}}, \citenamefont {{Coughlin}}, \citenamefont {{Morris}},
  \citenamefont {{Regimbau}}, \citenamefont {{Christensen}}, \citenamefont
  {{Kandhasamy}}, \citenamefont {{Mandic}}, \citenamefont {{Romano}},\ and\
  \citenamefont {{Thrane}}}]{2015PhRvD..92f3002M}%
  \BibitemOpen
  \bibfield  {author} {\bibinfo {author} {\bibfnamefont {D.}~\bibnamefont
  {{Meacher}}}, \bibinfo {author} {\bibfnamefont {M.}~\bibnamefont
  {{Coughlin}}}, \bibinfo {author} {\bibfnamefont {S.}~\bibnamefont
  {{Morris}}}, \bibinfo {author} {\bibfnamefont {T.}~\bibnamefont
  {{Regimbau}}}, \bibinfo {author} {\bibfnamefont {N.}~\bibnamefont
  {{Christensen}}}, \bibinfo {author} {\bibfnamefont {S.}~\bibnamefont
  {{Kandhasamy}}}, \bibinfo {author} {\bibfnamefont {V.}~\bibnamefont
  {{Mandic}}}, \bibinfo {author} {\bibfnamefont {J.~D.}\ \bibnamefont
  {{Romano}}},\ and\ \bibinfo {author} {\bibfnamefont {E.}~\bibnamefont
  {{Thrane}}},\ }\href {https://doi.org/10.1103/PhysRevD.92.063002} {\bibfield
  {journal} {\bibinfo  {journal} {Phys. Rev. D}\ }\textbf {\bibinfo {volume}
  {92}},\ \bibinfo {eid} {063002} (\bibinfo {year} {2015})},\ \Eprint
  {https://arxiv.org/abs/1506.06744} {arXiv:1506.06744 [astro-ph.HE]}
  \BibitemShut {NoStop}%
\bibitem [{\citenamefont {{Gupta, Ish et al.}}(2024)}]{network_sensi_main}%
  \BibitemOpen
  \bibfield  {author} {\bibinfo {author} {\bibnamefont {{Gupta, Ish et al.}}},\
  }\href {https://doi.org/10.1088/1361-6382/ad7b99} {\bibfield  {journal}
  {\bibinfo  {journal} {Classical and Quantum Gravity}\ }\textbf {\bibinfo
  {volume} {41}},\ \bibinfo {eid} {245001} (\bibinfo {year} {2024})},\ \Eprint
  {https://arxiv.org/abs/2307.10421} {arXiv:2307.10421 [gr-qc]} \BibitemShut
  {NoStop}%
\bibitem [{\citenamefont {{Reitze, David et al.}}(2019{\natexlab{a}})}]{CE}%
  \BibitemOpen
  \bibfield  {author} {\bibinfo {author} {\bibnamefont {{Reitze, David et
  al.}}},\ }in\ \href {https://doi.org/10.48550/arXiv.1907.04833} {\emph
  {\bibinfo {booktitle} {Bulletin of the American Astronomical Society}}},\
  Vol.~\bibinfo {volume} {51}\ (\bibinfo {year} {2019})\ p.~\bibinfo {pages}
  {35},\ \Eprint {https://arxiv.org/abs/1907.04833} {arXiv:1907.04833
  [astro-ph.IM]} \BibitemShut {NoStop}%
\bibitem [{\citenamefont {{Maggiore}}\ \emph {et~al.}(2024)\citenamefont
  {{Maggiore}}, \citenamefont {{Iacovelli}}, \citenamefont {{Belgacem}},
  \citenamefont {{Mancarella}},\ and\ \citenamefont
  {{Muttoni}}}]{Maggiore2024}%
  \BibitemOpen
  \bibfield  {author} {\bibinfo {author} {\bibfnamefont {M.}~\bibnamefont
  {{Maggiore}}}, \bibinfo {author} {\bibfnamefont {F.}~\bibnamefont
  {{Iacovelli}}}, \bibinfo {author} {\bibfnamefont {E.}~\bibnamefont
  {{Belgacem}}}, \bibinfo {author} {\bibfnamefont {M.}~\bibnamefont
  {{Mancarella}}},\ and\ \bibinfo {author} {\bibfnamefont {N.}~\bibnamefont
  {{Muttoni}}},\ }\href {https://doi.org/10.48550/arXiv.2411.05754} {\bibfield
  {journal} {\bibinfo  {journal} {arXiv e-prints}\ ,\ \bibinfo {eid}
  {arXiv:2411.05754}} (\bibinfo {year} {2024})},\ \Eprint
  {https://arxiv.org/abs/2411.05754} {arXiv:2411.05754 [gr-qc]} \BibitemShut
  {NoStop}%
\bibitem [{\citenamefont {{Gill}}\ \emph {et~al.}(2019)\citenamefont {{Gill}},
  \citenamefont {{Nathanail}},\ and\ \citenamefont {{Rezzolla}}}]{Gill2019}%
  \BibitemOpen
  \bibfield  {author} {\bibinfo {author} {\bibfnamefont {R.}~\bibnamefont
  {{Gill}}}, \bibinfo {author} {\bibfnamefont {A.}~\bibnamefont
  {{Nathanail}}},\ and\ \bibinfo {author} {\bibfnamefont {L.}~\bibnamefont
  {{Rezzolla}}},\ }\href {https://doi.org/10.3847/1538-4357/ab16da} {\bibfield
  {journal} {\bibinfo  {journal} {Astrophys. J.}\ }\textbf {\bibinfo {volume}
  {876}},\ \bibinfo {eid} {139} (\bibinfo {year} {2019})},\ \Eprint
  {https://arxiv.org/abs/1901.04138} {arXiv:1901.04138 [astro-ph.HE]}
  \BibitemShut {NoStop}%
\bibitem [{\citenamefont {{Murguia-Berthier}}\ \emph
  {et~al.}(2021)\citenamefont {{Murguia-Berthier}}, \citenamefont
  {{Ramirez-Ruiz}}, \citenamefont {{De Colle}}, \citenamefont {{Janiuk}},
  \citenamefont {{Rosswog}},\ and\ \citenamefont
  {{Lee}}}]{Murguia-Berthier2020}%
  \BibitemOpen
  \bibfield  {author} {\bibinfo {author} {\bibfnamefont {A.}~\bibnamefont
  {{Murguia-Berthier}}}, \bibinfo {author} {\bibfnamefont {E.}~\bibnamefont
  {{Ramirez-Ruiz}}}, \bibinfo {author} {\bibfnamefont {F.}~\bibnamefont {{De
  Colle}}}, \bibinfo {author} {\bibfnamefont {A.}~\bibnamefont {{Janiuk}}},
  \bibinfo {author} {\bibfnamefont {S.}~\bibnamefont {{Rosswog}}},\ and\
  \bibinfo {author} {\bibfnamefont {W.~H.}\ \bibnamefont {{Lee}}},\ }\href
  {https://doi.org/10.3847/1538-4357/abd08e} {\bibfield  {journal} {\bibinfo
  {journal} {Astrophys. J.}\ }\textbf {\bibinfo {volume} {908}},\ \bibinfo
  {eid} {152} (\bibinfo {year} {2021})},\ \Eprint
  {https://arxiv.org/abs/2007.12245} {arXiv:2007.12245 [astro-ph.HE]}
  \BibitemShut {NoStop}%
\bibitem [{\citenamefont {{{LIGO} Scientific
  Collaboration}}(2020)}]{2020ascl.soft12021L}%
  \BibitemOpen
  \bibfield  {author} {\bibinfo {author} {\bibnamefont {{{LIGO} Scientific
  Collaboration}}},\ }\href@noop {} {\bibinfo {title} {{LALSuite: LIGO
  Scientific Collaboration Algorithm Library Suite}}},\ \bibinfo {howpublished}
  {Astrophysics Source Code Library, record ascl:2012.021} (\bibinfo {year}
  {2020})\BibitemShut {NoStop}%
\bibitem [{\citenamefont {{Husa}}\ \emph {et~al.}(2016)\citenamefont {{Husa}},
  \citenamefont {{Khan}}, \citenamefont {{Hannam}}, \citenamefont
  {{P{\"u}rrer}}, \citenamefont {{Ohme}}, \citenamefont {{Forteza}},\ and\
  \citenamefont {{Boh{\'e}}}}]{PhenomD1}%
  \BibitemOpen
  \bibfield  {author} {\bibinfo {author} {\bibfnamefont {S.}~\bibnamefont
  {{Husa}}}, \bibinfo {author} {\bibfnamefont {S.}~\bibnamefont {{Khan}}},
  \bibinfo {author} {\bibfnamefont {M.}~\bibnamefont {{Hannam}}}, \bibinfo
  {author} {\bibfnamefont {M.}~\bibnamefont {{P{\"u}rrer}}}, \bibinfo {author}
  {\bibfnamefont {F.}~\bibnamefont {{Ohme}}}, \bibinfo {author} {\bibfnamefont
  {X.~J.}\ \bibnamefont {{Forteza}}},\ and\ \bibinfo {author} {\bibfnamefont
  {A.}~\bibnamefont {{Boh{\'e}}}},\ }\href
  {https://doi.org/10.1103/PhysRevD.93.044006} {\bibfield  {journal} {\bibinfo
  {journal} {Phys. Rev. D}\ }\textbf {\bibinfo {volume} {93}},\ \bibinfo {eid}
  {044006} (\bibinfo {year} {2016})},\ \Eprint
  {https://arxiv.org/abs/1508.07250} {arXiv:1508.07250 [gr-qc]} \BibitemShut
  {NoStop}%
\bibitem [{\citenamefont {{Khan}}\ \emph {et~al.}(2016)\citenamefont {{Khan}},
  \citenamefont {{Husa}}, \citenamefont {{Hannam}}, \citenamefont {{Ohme}},
  \citenamefont {{P{\"u}rrer}}, \citenamefont {{Forteza}},\ and\ \citenamefont
  {{Boh{\'e}}}}]{PhenomD2}%
  \BibitemOpen
  \bibfield  {author} {\bibinfo {author} {\bibfnamefont {S.}~\bibnamefont
  {{Khan}}}, \bibinfo {author} {\bibfnamefont {S.}~\bibnamefont {{Husa}}},
  \bibinfo {author} {\bibfnamefont {M.}~\bibnamefont {{Hannam}}}, \bibinfo
  {author} {\bibfnamefont {F.}~\bibnamefont {{Ohme}}}, \bibinfo {author}
  {\bibfnamefont {M.}~\bibnamefont {{P{\"u}rrer}}}, \bibinfo {author}
  {\bibfnamefont {X.~J.}\ \bibnamefont {{Forteza}}},\ and\ \bibinfo {author}
  {\bibfnamefont {A.}~\bibnamefont {{Boh{\'e}}}},\ }\href
  {https://doi.org/10.1103/PhysRevD.93.044007} {\bibfield  {journal} {\bibinfo
  {journal} {Phys. Rev. D}\ }\textbf {\bibinfo {volume} {93}},\ \bibinfo {eid}
  {044007} (\bibinfo {year} {2016})},\ \Eprint
  {https://arxiv.org/abs/1508.07253} {arXiv:1508.07253 [gr-qc]} \BibitemShut
  {NoStop}%
\bibitem [{\citenamefont {{Koeppel}}\ \emph {et~al.}(2019)\citenamefont
  {{Koeppel}}, \citenamefont {{Bovard}},\ and\ \citenamefont
  {{Rezzolla}}}]{Koeppel2019}%
  \BibitemOpen
  \bibfield  {author} {\bibinfo {author} {\bibfnamefont {S.}~\bibnamefont
  {{Koeppel}}}, \bibinfo {author} {\bibfnamefont {L.}~\bibnamefont
  {{Bovard}}},\ and\ \bibinfo {author} {\bibfnamefont {L.}~\bibnamefont
  {{Rezzolla}}},\ }\href {https://doi.org/10.3847/2041-8213/ab0210} {\bibfield
  {journal} {\bibinfo  {journal} {Astrophys. J. Lett.}\ }\textbf {\bibinfo
  {volume} {872}},\ \bibinfo {eid} {L16} (\bibinfo {year} {2019})},\ \Eprint
  {https://arxiv.org/abs/1901.09977} {arXiv:1901.09977 [gr-qc]} \BibitemShut
  {NoStop}%
\bibitem [{\citenamefont {{Bauswein}}\ \emph {et~al.}(2013)\citenamefont
  {{Bauswein}}, \citenamefont {{Baumgarte}},\ and\ \citenamefont
  {{Janka}}}]{Bauswein2013}%
  \BibitemOpen
  \bibfield  {author} {\bibinfo {author} {\bibfnamefont {A.}~\bibnamefont
  {{Bauswein}}}, \bibinfo {author} {\bibfnamefont {T.~W.}\ \bibnamefont
  {{Baumgarte}}},\ and\ \bibinfo {author} {\bibfnamefont {H.-T.}\ \bibnamefont
  {{Janka}}},\ }\href {https://doi.org/10.1103/PhysRevLett.111.131101}
  {\bibfield  {journal} {\bibinfo  {journal} {Phys. Rev. Lett.}\ }\textbf
  {\bibinfo {volume} {111}},\ \bibinfo {eid} {131101} (\bibinfo {year}
  {2013})},\ \Eprint {https://arxiv.org/abs/1307.5191} {arXiv:1307.5191
  [astro-ph.SR]} \BibitemShut {NoStop}%
\bibitem [{\citenamefont {{Tootle}}\ \emph {et~al.}(2021)\citenamefont
  {{Tootle}}, \citenamefont {{Papenfort}}, \citenamefont {{Most}},\ and\
  \citenamefont {{Rezzolla}}}]{Tootle2021}%
  \BibitemOpen
  \bibfield  {author} {\bibinfo {author} {\bibfnamefont {S.~D.}\ \bibnamefont
  {{Tootle}}}, \bibinfo {author} {\bibfnamefont {L.~J.}\ \bibnamefont
  {{Papenfort}}}, \bibinfo {author} {\bibfnamefont {E.~R.}\ \bibnamefont
  {{Most}}},\ and\ \bibinfo {author} {\bibfnamefont {L.}~\bibnamefont
  {{Rezzolla}}},\ }\href {https://doi.org/10.3847/2041-8213/ac350d} {\bibfield
  {journal} {\bibinfo  {journal} {Astrophys. J. Lett.}\ }\textbf {\bibinfo
  {volume} {922}},\ \bibinfo {eid} {L19} (\bibinfo {year} {2021})},\ \Eprint
  {https://arxiv.org/abs/2109.00940} {arXiv:2109.00940 [gr-qc]} \BibitemShut
  {NoStop}%
\bibitem [{\citenamefont {{K{\"o}lsch}}\ \emph {et~al.}(2022)\citenamefont
  {{K{\"o}lsch}}, \citenamefont {{Dietrich}}, \citenamefont {{Ujevic}},\ and\
  \citenamefont {{Br{\"u}gmann}}}]{Koelsch2021}%
  \BibitemOpen
  \bibfield  {author} {\bibinfo {author} {\bibfnamefont {M.}~\bibnamefont
  {{K{\"o}lsch}}}, \bibinfo {author} {\bibfnamefont {T.}~\bibnamefont
  {{Dietrich}}}, \bibinfo {author} {\bibfnamefont {M.}~\bibnamefont
  {{Ujevic}}},\ and\ \bibinfo {author} {\bibfnamefont {B.}~\bibnamefont
  {{Br{\"u}gmann}}},\ }\href {https://doi.org/10.1103/PhysRevD.106.044026}
  {\bibfield  {journal} {\bibinfo  {journal} {Phys. Rev. D}\ }\textbf {\bibinfo
  {volume} {106}},\ \bibinfo {eid} {044026} (\bibinfo {year} {2022})},\ \Eprint
  {https://arxiv.org/abs/2112.11851} {arXiv:2112.11851 [gr-qc]} \BibitemShut
  {NoStop}%
\bibitem [{\citenamefont {{Kashyap}}\ \emph {et~al.}(2022)\citenamefont
  {{Kashyap}}, \citenamefont {{Das}}, \citenamefont {{Radice}}, \citenamefont
  {{Padamata}}, \citenamefont {{Prakash}}, \citenamefont {{Logoteta}},
  \citenamefont {{Perego}}, \citenamefont {{Godzieba}}, \citenamefont
  {{Bernuzzi}}, \citenamefont {{Bombaci}}, \citenamefont {{Fattoyev}},
  \citenamefont {{Reed}},\ and\ \citenamefont {{Schneider}}}]{Kashyap:2021wzs}%
  \BibitemOpen
  \bibfield  {author} {\bibinfo {author} {\bibfnamefont {R.}~\bibnamefont
  {{Kashyap}}}, \bibinfo {author} {\bibfnamefont {A.}~\bibnamefont {{Das}}},
  \bibinfo {author} {\bibfnamefont {D.}~\bibnamefont {{Radice}}}, \bibinfo
  {author} {\bibfnamefont {S.}~\bibnamefont {{Padamata}}}, \bibinfo {author}
  {\bibfnamefont {A.}~\bibnamefont {{Prakash}}}, \bibinfo {author}
  {\bibfnamefont {D.}~\bibnamefont {{Logoteta}}}, \bibinfo {author}
  {\bibfnamefont {A.}~\bibnamefont {{Perego}}}, \bibinfo {author}
  {\bibfnamefont {D.~A.}\ \bibnamefont {{Godzieba}}}, \bibinfo {author}
  {\bibfnamefont {S.}~\bibnamefont {{Bernuzzi}}}, \bibinfo {author}
  {\bibfnamefont {I.}~\bibnamefont {{Bombaci}}}, \bibinfo {author}
  {\bibfnamefont {F.~J.}\ \bibnamefont {{Fattoyev}}}, \bibinfo {author}
  {\bibfnamefont {B.~T.}\ \bibnamefont {{Reed}}},\ and\ \bibinfo {author}
  {\bibfnamefont {A.~d.~S.}\ \bibnamefont {{Schneider}}},\ }\href
  {https://doi.org/10.1103/PhysRevD.105.103022} {\bibfield  {journal} {\bibinfo
   {journal} {Phys. Rev. D}\ }\textbf {\bibinfo {volume} {105}},\ \bibinfo
  {eid} {103022} (\bibinfo {year} {2022})},\ \Eprint
  {https://arxiv.org/abs/2111.05183} {arXiv:2111.05183 [astro-ph.HE]}
  \BibitemShut {NoStop}%
\bibitem [{\citenamefont {{Paschalidis}}(2017)}]{Paschalidis2016}%
  \BibitemOpen
  \bibfield  {author} {\bibinfo {author} {\bibfnamefont {V.}~\bibnamefont
  {{Paschalidis}}},\ }\href {https://doi.org/10.1088/1361-6382/aa61ce}
  {\bibfield  {journal} {\bibinfo  {journal} {Classical and Quantum Gravity}\
  }\textbf {\bibinfo {volume} {34}},\ \bibinfo {eid} {084002} (\bibinfo {year}
  {2017})},\ \Eprint {https://arxiv.org/abs/1611.01519} {arXiv:1611.01519
  [astro-ph.HE]} \BibitemShut {NoStop}%
\bibitem [{\citenamefont {{Takami}}\ \emph
  {et~al.}(2014{\natexlab{b}})\citenamefont {{Takami}}, \citenamefont
  {{Rezzolla}},\ and\ \citenamefont {{Baiotti}}}]{Takami2014}%
  \BibitemOpen
  \bibfield  {author} {\bibinfo {author} {\bibfnamefont {K.}~\bibnamefont
  {{Takami}}}, \bibinfo {author} {\bibfnamefont {L.}~\bibnamefont
  {{Rezzolla}}},\ and\ \bibinfo {author} {\bibfnamefont {L.}~\bibnamefont
  {{Baiotti}}},\ }\href {https://doi.org/10.1103/PhysRevLett.113.091104}
  {\bibfield  {journal} {\bibinfo  {journal} {\prl}\ }\textbf {\bibinfo
  {volume} {113}},\ \bibinfo {eid} {091104} (\bibinfo {year}
  {2014}{\natexlab{b}})},\ \Eprint {https://arxiv.org/abs/1403.5672}
  {arXiv:1403.5672 [gr-qc]} \BibitemShut {NoStop}%
\bibitem [{\citenamefont {{Takami}}\ \emph
  {et~al.}(2015{\natexlab{b}})\citenamefont {{Takami}}, \citenamefont
  {{Rezzolla}},\ and\ \citenamefont {{Baiotti}}}]{Takami2015}%
  \BibitemOpen
  \bibfield  {author} {\bibinfo {author} {\bibfnamefont {K.}~\bibnamefont
  {{Takami}}}, \bibinfo {author} {\bibfnamefont {L.}~\bibnamefont
  {{Rezzolla}}},\ and\ \bibinfo {author} {\bibfnamefont {L.}~\bibnamefont
  {{Baiotti}}},\ }\href {https://doi.org/10.1103/PhysRevD.91.064001} {\bibfield
   {journal} {\bibinfo  {journal} {\prd}\ }\textbf {\bibinfo {volume} {91}},\
  \bibinfo {eid} {064001} (\bibinfo {year} {2015}{\natexlab{b}})},\ \Eprint
  {https://arxiv.org/abs/1412.3240} {arXiv:1412.3240 [gr-qc]} \BibitemShut
  {NoStop}%
\bibitem [{\citenamefont {{Bauswein}}\ and\ \citenamefont
  {{Stergioulas}}(2015{\natexlab{b}})}]{Bauswein2015}%
  \BibitemOpen
  \bibfield  {author} {\bibinfo {author} {\bibfnamefont {A.}~\bibnamefont
  {{Bauswein}}}\ and\ \bibinfo {author} {\bibfnamefont {N.}~\bibnamefont
  {{Stergioulas}}},\ }\href {https://doi.org/10.1103/PhysRevD.91.124056}
  {\bibfield  {journal} {\bibinfo  {journal} {\prd}\ }\textbf {\bibinfo
  {volume} {91}},\ \bibinfo {eid} {124056} (\bibinfo {year}
  {2015}{\natexlab{b}})},\ \Eprint {https://arxiv.org/abs/1502.03176}
  {arXiv:1502.03176 [astro-ph.SR]} \BibitemShut {NoStop}%
\bibitem [{\citenamefont {{Farrow}}\ \emph {et~al.}(2019)\citenamefont
  {{Farrow}}, \citenamefont {{Zhu}},\ and\ \citenamefont
  {{Thrane}}}]{farrow_mass_2019}%
  \BibitemOpen
  \bibfield  {author} {\bibinfo {author} {\bibfnamefont {N.}~\bibnamefont
  {{Farrow}}}, \bibinfo {author} {\bibfnamefont {X.-J.}\ \bibnamefont
  {{Zhu}}},\ and\ \bibinfo {author} {\bibfnamefont {E.}~\bibnamefont
  {{Thrane}}},\ }\href {https://doi.org/10.3847/1538-4357/ab12e3} {\bibfield
  {journal} {\bibinfo  {journal} {Astrophys. J.}\ }\textbf {\bibinfo {volume}
  {876}},\ \bibinfo {eid} {18} (\bibinfo {year} {2019})},\ \Eprint
  {https://arxiv.org/abs/1902.03300} {arXiv:1902.03300 [astro-ph.HE]}
  \BibitemShut {NoStop}%
\bibitem [{\citenamefont {{Romero-Shaw}}\ \emph {et~al.}(2020)\citenamefont
  {{Romero-Shaw}}, \citenamefont {{Farrow}}, \citenamefont {{Stevenson}},
  \citenamefont {{Thrane}},\ and\ \citenamefont {{Zhu}}}]{2020MNRAS.496L..64R}%
  \BibitemOpen
  \bibfield  {author} {\bibinfo {author} {\bibfnamefont {I.~M.}\ \bibnamefont
  {{Romero-Shaw}}}, \bibinfo {author} {\bibfnamefont {N.}~\bibnamefont
  {{Farrow}}}, \bibinfo {author} {\bibfnamefont {S.}~\bibnamefont
  {{Stevenson}}}, \bibinfo {author} {\bibfnamefont {E.}~\bibnamefont
  {{Thrane}}},\ and\ \bibinfo {author} {\bibfnamefont {X.-J.}\ \bibnamefont
  {{Zhu}}},\ }\href {https://doi.org/10.1093/mnrasl/slaa084} {\bibfield
  {journal} {\bibinfo  {journal} {Mon. Not. R. Astron. Soc.}\ }\textbf
  {\bibinfo {volume} {496}},\ \bibinfo {pages} {L64} (\bibinfo {year}
  {2020})},\ \Eprint {https://arxiv.org/abs/2001.06492} {arXiv:2001.06492
  [astro-ph.HE]} \BibitemShut {NoStop}%
\bibitem [{\citenamefont {{Abbott, R. et
  al.}}(2020)}]{the_ligo_scientific_collaboration_gw190814_2020}%
  \BibitemOpen
  \bibfield  {author} {\bibinfo {author} {\bibnamefont {{Abbott, R. et al.}}},\
  }\href {https://doi.org/10.3847/2041-8213/ab960f} {\bibfield  {journal}
  {\bibinfo  {journal} {Astrophys. J. Lett.}\ }\textbf {\bibinfo {volume}
  {896}},\ \bibinfo {eid} {L44} (\bibinfo {year} {2020})},\ \Eprint
  {https://arxiv.org/abs/2006.12611} {arXiv:2006.12611 [astro-ph.HE]}
  \BibitemShut {NoStop}%
\bibitem [{\citenamefont {{Allen}}(1997)}]{1997rggr.conf..373A}%
  \BibitemOpen
  \bibfield  {author} {\bibinfo {author} {\bibfnamefont {B.}~\bibnamefont
  {{Allen}}},\ }in\ \href {https://doi.org/10.48550/arXiv.gr-qc/9604033} {\emph
  {\bibinfo {booktitle} {Relativistic Gravitation and Gravitational
  Radiation}}},\ \bibinfo {editor} {edited by\ \bibinfo {editor} {\bibfnamefont
  {J.-A.}\ \bibnamefont {{Marck}}}\ and\ \bibinfo {editor} {\bibfnamefont
  {J.-P.}\ \bibnamefont {{Lasota}}}}\ (\bibinfo {year} {1997})\ pp.\ \bibinfo
  {pages} {373--417},\ \Eprint {https://arxiv.org/abs/gr-qc/9604033}
  {arXiv:gr-qc/9604033 [gr-qc]} \BibitemShut {NoStop}%
\bibitem [{\citenamefont {{Christensen}}(1992)}]{1992PhRvD..46.5250C}%
  \BibitemOpen
  \bibfield  {author} {\bibinfo {author} {\bibfnamefont {N.}~\bibnamefont
  {{Christensen}}},\ }\href {https://doi.org/10.1103/PhysRevD.46.5250}
  {\bibfield  {journal} {\bibinfo  {journal} {Phys. Rev. D}\ }\textbf {\bibinfo
  {volume} {46}},\ \bibinfo {pages} {5250} (\bibinfo {year}
  {1992})}\BibitemShut {NoStop}%
\bibitem [{\citenamefont {{Thrane}}\ and\ \citenamefont
  {{Romano}}(2013)}]{thrane_sensitivity_2013}%
  \BibitemOpen
  \bibfield  {author} {\bibinfo {author} {\bibfnamefont {E.}~\bibnamefont
  {{Thrane}}}\ and\ \bibinfo {author} {\bibfnamefont {J.~D.}\ \bibnamefont
  {{Romano}}},\ }\href {https://doi.org/10.1103/PhysRevD.88.124032} {\bibfield
  {journal} {\bibinfo  {journal} {Phys. Rev. D}\ }\textbf {\bibinfo {volume}
  {88}},\ \bibinfo {eid} {124032} (\bibinfo {year} {2013})},\ \Eprint
  {https://arxiv.org/abs/1310.5300} {arXiv:1310.5300 [astro-ph.IM]}
  \BibitemShut {NoStop}%
\bibitem [{\citenamefont {{Flanagan}}(1993)}]{1993PhRvD..48.2389F}%
  \BibitemOpen
  \bibfield  {author} {\bibinfo {author} {\bibfnamefont {E.~E.}\ \bibnamefont
  {{Flanagan}}},\ }\href {https://doi.org/10.1103/PhysRevD.48.2389} {\bibfield
  {journal} {\bibinfo  {journal} {Phys. Rev. D}\ }\textbf {\bibinfo {volume}
  {48}},\ \bibinfo {pages} {2389} (\bibinfo {year} {1993})},\ \Eprint
  {https://arxiv.org/abs/astro-ph/9305029} {arXiv:astro-ph/9305029 [astro-ph]}
  \BibitemShut {NoStop}%
\bibitem [{\citenamefont {{Reitze, David et
  al.}}(2019{\natexlab{b}})}]{CE_science_case}%
  \BibitemOpen
  \bibfield  {author} {\bibinfo {author} {\bibnamefont {{Reitze, David et
  al.}}},\ }in\ \href {https://doi.org/10.48550/arXiv.1907.04833} {\emph
  {\bibinfo {booktitle} {Bulletin of the American Astronomical Society}}},\
  Vol.~\bibinfo {volume} {51}\ (\bibinfo {year} {2019})\ p.~\bibinfo {pages}
  {35},\ \Eprint {https://arxiv.org/abs/1907.04833} {arXiv:1907.04833
  [astro-ph.IM]} \BibitemShut {NoStop}%
\bibitem [{\citenamefont {{Janssens}}\ \emph {et~al.}(2022)\citenamefont
  {{Janssens}}, \citenamefont {{Boileau}}, \citenamefont {{Christensen}},
  \citenamefont {{Badaracco}},\ and\ \citenamefont {{van
  Remortel}}}]{2022PhRvD.106d2008J}%
  \BibitemOpen
  \bibfield  {author} {\bibinfo {author} {\bibfnamefont {K.}~\bibnamefont
  {{Janssens}}}, \bibinfo {author} {\bibfnamefont {G.}~\bibnamefont
  {{Boileau}}}, \bibinfo {author} {\bibfnamefont {N.}~\bibnamefont
  {{Christensen}}}, \bibinfo {author} {\bibfnamefont {F.}~\bibnamefont
  {{Badaracco}}},\ and\ \bibinfo {author} {\bibfnamefont {N.}~\bibnamefont
  {{van Remortel}}},\ }\href {https://doi.org/10.1103/PhysRevD.106.042008}
  {\bibfield  {journal} {\bibinfo  {journal} {Phys. Rev. D}\ }\textbf {\bibinfo
  {volume} {106}},\ \bibinfo {eid} {042008} (\bibinfo {year} {2022})},\ \Eprint
  {https://arxiv.org/abs/2206.06809} {arXiv:2206.06809 [astro-ph.IM]}
  \BibitemShut {NoStop}%
\end{thebibliography}%

\end{document}